\documentclass[preprint showpacs,preprintnumbers,amsmath,amssymb]{revtex4}
\usepackage{amssymb}
\usepackage{longtable}
\usepackage{xcolor}
\usepackage{graphicx}
\usepackage{dcolumn}
\usepackage{bm}
\usepackage{multirow}
\usepackage{natbib}
\usepackage{amsmath}	
\usepackage{color}
\usepackage{cases} 

\newcommand{\Vmat}{{\boldmath $\cal V$}}

\begin{document}
\preprint{APS/123-QED}

\title{Electron-induced excitation, recombination and dissociation of molecular ions initiating the formation of complex organic molecules}

\author{J. Zs. Mezei$^{1}$}
\email[]{mezei.zsolt@atomki.hu}
\author{K. Chakrabarti$^{2}$}
\author{M. D. Ep\'ee Ep\'ee$^{3}$}
\author{O. Motapon$^{3}$}
\author{C. H. Yuen$^{4}$}
\author{M. A. Ayouz$^{5}$}
\author{N. Douguet$^{4}$}
\email{nicolas.douguet@ucf.edu}
\author{S. Fonseca dos Santos$^6$}
\email{sfonseca@rollins.edu}
\author{V. Kokoouline$^4$}
\email{slavako@ucf.edu}
\author{I. F. Schneider$^{7,8}$}
\email{ioan.schneider@univ-lehavre.fr}
\affiliation{$^{1}$Institute for Nuclear Research (ATOMKI), H-4001 Debrecen, Hungary}%
\affiliation{$^{2}$Department of Mathematics, Scottish Church College, 700006 Kolkata, India}
\affiliation{$^{3}$UFD Math\'ematiques, Informatique Appliqu\'ee et Physique Fondamentale, University of Douala, PO Box 24157 Douala, Cameroon}
\affiliation{$^{4}$Department of Physics, University of Central Florida, Orlando, FL, U.S.A.}
\affiliation{$^{5}$Laboratoire G\'enie des Proc\'ed\'es et Mat\'eriaux CNRS-EA4038, CentralSup\'elec, Universit\'e Paris-Saclay, F-91190 Gif-sur-Yvette, France}
\affiliation{$^{6}$Rollins College, Winter Park, FL, U.S.A.}
\affiliation{$^{7}$Laboratoire Ondes \& Milieux Complexes CNRS-UMR6294, Universit\'e du Havre, Normandie Universit\'e, F-76058 Le Havre, France}
\affiliation{$^{8}$Laboratoire Aim\'e Cotton CNRS-FRE2038, Universit\'e Paris-Saclay, F-91405 Orsay, France}
\date{\today}

\begin{abstract}
We review the study of dissociative recombination and ro-vibrational excitation of 
 diatomic and small polyatomic molecular ions initiating complex organic molecules formation.
In particular, we show how Multichannel Quantum Defect Theory (MQDT)
and R-matrix methods are used to compute cross sections and rate coefficients for cations in well defined ro-vibrational levels of the ground electronic state, from sub-meV up to few eV collision energies.

The most recent MQDT results are compared with either other theoretical data, or with measured data obtained in storage-ring experiments.
\end{abstract}


\maketitle

\section{\label{sec:intro}Introduction}

The cold ionized media of astrophysical interest, namely the interstellar molecular clouds, the supernovae, the planetary atmospheres, etc., are the seat of an extremely rich chemical physics, due to the presence of numerous atomic and molecular species - neutral or ionized - photons, low energy electrons and cosmic rays.

There is a variety of processes that can lead to the formation/destruction of molecules in the interstellar medium (ISM), but these can be separated into two broad classes: reactions that occur in the gas phase and reactions that occur on the surfaces of small grains prevalent throughout the interstellar medium. 

The reactions taking place in the gas phase can be further divided~\citep{tielens2005} into {\it bond-forming} processes, including radiative association, which link atoms 
and molecules into more complex species, i.e. complex organic molecules (COMs) or polycyclic aromatic hydrocarbons (PAHs), and {\it bond-destruction} processes, such as photoionization, photodissociation, atomic- and molecular-induced collisional dissociation 
and electron-impact dissociative recombination, which result into stable and metastable smaller species and/or radicals. 

Finally, the {\it bond-rearrangement} reactions - ion-molecule charge-transfer reactions, and neutral-neutral reactions - transfer parts of one co-reactant to another one. 

An important case is the carbon chemistry of the diffuse~\citep{dopita2003} and/or dense~\citep{herbst1973} ISM, which starts 
from the carbon atom and results through the bond-forming/rearrangement processes to complex molecules like HCO, HOC, HCN, CH$_3$OH, 
HCOOCH$_3$, CH$_3$NH$_2$, as, e.g., in the following chain of reactions~\citep{herbst2008}:

\begin{eqnarray}
\label{com1}\mbox{C}+\mbox{H}_3^+ &\rightarrow& \mbox{CH}^++\mbox{H}_2\rightarrow \mbox{CH}^+_2+\mbox{H}\\
\label{com2}\mbox{CH}^+_2+\mbox{H}_2 &\rightarrow& \mbox{CH}^+_3+\mbox{H}\\
\label{com3}\mbox{CH}^+_3+\mbox{NH}_3 &\rightarrow& \mbox{CH}_3\mbox{NH}_3^++h\nu\\
\label{com4}\mbox{CH}_3\mbox{NH}_3^++e^-&\rightarrow& \mbox{CH}_3\mbox{NH}_2+\mbox{H}
\end{eqnarray}
 
\noindent 
More examples can be found in figs. 15, 16, 19, 20 and 22 of Ref.~\citep{tielens2013} where many other pathways for COMs formations are proposed. 

In essentially all of these formation pathways relevant for the ISM environments, where the temperature and pressure are very low, carbon containing diatomic or polyatomic molecular cations CX$^+$ are involved.   The abundance of carbon-containing molecules is partly driven by the electron induced dissociative recombination (DR):

\begin{equation}
\label{eq:DR} \mbox{CX}^{+}(N_{i}^{+},v_{i}^{+}) + e^{-} \longrightarrow
\mbox{CX}^{*},\mbox{CX}^{**} \longrightarrow
\mbox{C}^{*} + \mbox{X}\,\,\,\,\text{or}\,\,\,\,\mbox{C} + \mbox{X}^{*}
,
\end{equation}

\noindent where X stands for atomic or molecular species containing hydrogen, oxygen, fluorine, etc., $ N_{i}^{+} $ and $ v_{i}^{+} $ denote respectively, the initial ro-vibrational levels of the  molecular ion in its electronic ground-state, CX$^{**}$ stands for dissociative autoionizing - often doubly-excited - states, and CX$^*$ for bound excited states belonging to Rydberg series of the neutral. 

Meanwhile, within the same reactive collision, this ion destruction process competes with transitions between the ro-vibrational states of the target:

 \begin{equation}
\label{eq:SEC} 
\mbox{CX}^{+}(N_{i}^{+}, v_{i}^{+}) +
e^{-}(\varepsilon) \longrightarrow
\mbox{CX}^{+}(N_{f}^{+},v_{f}^{+}) +
e^{-}({\varepsilon}^{'})
,
\end{equation}
\noindent where 
$N_{f}^{+}$ and $ v_f^+ $ are its final rotational and vibrational quantum numbers, and $\varepsilon$/$\varepsilon'$ the initial/final energy of the incident electron.

 In the often non-equilibrium and cold environments of astrochemical interest, the qualitative and quantitative understanding of the formation of COMs are critically based on the precise knowledge of state-to-state cross sections and/or rate coefficients of the electron induced dissociation and/or excitation of molecular cations.

The present paper is divided into two parts. 

The first one deals with the diatomic systems. The multichannel quantum defect theory (MQDT) is used to calculate state-to-state DR cross sections and rate coefficients for molecular cations like CH$^+$, CO$^+$ and CF$^+$. The detailed presentation of the method is followed by the results with special focus on the driving mechanisms.

The second part presents our results obtained for polyatomic systems. The MQDT and the normal mode approach combined with the R-matrix theory is applied for calculating DR cross sections for  H$_3$O$^+$, HCO$^+$, CH$_3^+$ and CH$_2$NH$_2^+$. 

The paper ends with conclusions and future plans.

\section{Multichannel quantum defect theory (MQDT) of electron/diatomic molecular cation collisions}

We currently use an MQDT-type method to study the electron-impact collision processes given by Eqs.~(\ref{eq:DR}) and (\ref{eq:SEC}). These processes involve {\it ionization} channels, describing the scattering of an electron on the molecular ion, and {\it dissociation} channels, accounting for atom-atom scattering. The mixing of these channels results in quantum interference of the {\it direct} mechanism - in which the capture takes place into a dissociative
state of the neutral system (CX$^{**}$) - and the {\it indirect} one -  in which the capture occurs \textit{via} a Rydberg state of the molecule CX$^{*}$,   predissociated by the CX$^{**}$ state.  
  The direct mechanism dominates the reactive collisions in the cases of favorable crossings (in the sense of the Franck-Condon principle) between the potential energy curves of the dissociative states and that of the target ion - CO$^+$ and CF$^+$ - and is exceeded by the indirect one otherwise - CH$^+$, as shown below.
In both mechanisms the autoionization is in competition with the predissociation, and leads, through the  reaction (\ref{eq:SEC}), to {\it super-elastic collision} 
(SEC,  $\varepsilon'>\varepsilon$),
{\it elastic collision} (EC, $\varepsilon'=\varepsilon$) and {\it inelastic collision} (IC, $\varepsilon'<\varepsilon$).

A detailed description of our theoretical approach has been given in previous studies on different diatomic systems, including the carbon-containing-ones~\citep{annick1980,carata2000,novotny2009,wafeu2011,
motapon2014,mezei2015,faure2017,moulane2018,chakrabarti2018}.  The main ideas and steps are recalled below for the three standard situations ($a$, $b$, $c$) encountered. 
This is performed in the order of the accuracy in predicting the cross section, 
from the very fine modeling of the rotational (a) and/or vibrational (a, b) resonances associated to the temporary capture into 
singly-excited Rydberg states, to that of the broad resonances associated to the capture into doubly-excited states (c). 
Whereas the major relevant details are provided for the $a$ case taken as reference, we outline  either the simplifications or the 
extensions in the situations  $b$ and $c$ with respect to the former. 

\vspace{.5cm}
\noindent
{\it
a) The account of {\em rotational} and {\em vibrational} structure and interactions for the target ion's  
{\em ground} electronic state and for the neutral's relevant electronic states~\citep{motapon2014}:  
}

The major steps in this case are the following:

\begin{enumerate}
\item {\it Building the interaction matrix} \Vmat :\\
Within a quasi-diabatic representation of the CX states, and for a given set of 
conserved quantum numbers  of the neutral system - $\Lambda$ (projection of the electronic angular momentum 
on the internuclear axis)  , $S$ (total electronic spin), $N$ (total rotational quantum number) - the interaction matrix is based on the 
couplings between 
\textit{ionisation} channels - 
	associated to
	the ro-vibrational levels $N^+v^+$ of the cation  
	and 
	to the orbital quantum number $l$ of the incident/Rydberg electron - 
and 
\textit{dissociation} channels $d_j$. The structure of the interaction matrix \Vmat introduced in step 1 is, in block form:\\
\begin{equation}\label{Vmat_1core_rot}
\boldsymbol{\mathcal{V}} = \left( \begin{array}{cc}
\ 0 & \boldsymbol{\mathcal{V}}_{{\bar d}~\overline {N_{c}v_{c}}} \\
\boldsymbol{\mathcal{V}}_{\overline{N_{c}v_{c}}~\bar d} & 0 \\
\end{array} \right),
\end{equation}
\noindent where the collective indices $\bar d$ and $\overline {N_{c}v_{c}}$ span the ensembles of all individual indices $d_j$ and $N^{+}v^{+}$ which respectively label dissociation channels and ionisation channels, the latter ones built on the ground electronic state of the ion - also called ground core, and labelled $c$. The only non-vanishing matrix elements organized in the non-diagonal blocks express the {\it Rydberg-valence} interaction.

\item {\it Computation of the reaction matrix} $\boldsymbol{\mathcal{K}}$:\\ 
Given $\boldsymbol{H_0}$ the Hamiltonian of the molecular system under study in which the Rydberg-valence interaction is neglected, we adopt  the second-order perturbative solution
for the Lippman-Schwinger integral equation \citep{ngassam2003}, 
written in operator form as:
\begin{equation}\label{eq:solveK}
\boldsymbol{\mathcal{K}}= \boldsymbol{\mathcal{V}} + \boldsymbol{\mathcal{V}}{\frac{1}{E-\boldsymbol{H_0}}}\boldsymbol{\mathcal{V}}.
\end{equation}

\item {\it Diagonalization of the reaction matrix,} 
yielding the eigenvectors and eigenvalues used to build the eigenchannel wave functions.

\item {\it Frame transformation} from 
{the Born-Oppenheimer (short range) representation, characterized by $N,v$ and $\Lambda$ quantum numbers,
valid for small electron-ion and nucleus-nucleus distances, 
to
the close-coupling (long-range) representation, characterized by  $N^+, v^+,\Lambda^+$ (for the ion) and $l$ 
(orbital quantum number of the incident/Rydberg electron), valid for both large distances.
This frame transformation relies on the}
quantum defects $\mu_{l}^{\Lambda}(R)$ describing the relevant Rydberg series built on the ionic core, and on the eigenvectors and eigenvalues of the K-matrix.

\item {\it Building of the {\it generalized} scattering 
matrix $\boldsymbol{{X}}$}, 
based on the frame-transformation coefficients, 
this matrix  being organised in blocks associated to energetically open and/or closed ($O$ and/or $C$ respectively) channels ("C" for "closed" to be distinguished from "c" for "core"):
\begin{equation}
\boldsymbol{{X}}=
 \left(\begin{array}{cc} \boldsymbol{X_{OO}} & \boldsymbol{X_{OC}}\\
                   \boldsymbol{X_{CO}} & \boldsymbol{X_{CC}} \end{array} \right)
.
\end{equation}

\item {\it Building  of the {\it physical} scattering matrix $\boldsymbol{\mathcal{S}}$}: 
\begin{equation}\label{eq:solve3}
\boldsymbol{S}=\boldsymbol{X_{OO}}-\boldsymbol{X_{OC}}\frac{1}{\boldsymbol{X_{CC}}-\exp(-i2\pi\boldsymbol{ \nu})}\boldsymbol{X_{CO}}
,
\end{equation}
\noindent
The first term in Eq.~({\ref{eq:solve3}}) is restricted to the open channels, resulting in the {\it direct} mechanism, and the second takes into account their mixing with the closed ones, resulting in the {\it total}, i.e. direct and indirect mechanism,  the denominator being responsible for the resonant patterns in the shape of the cross section
\citep{seaton1983}. Here the matrix $\exp(-i2\pi\boldsymbol{ \nu})$ is diagonal and relies on the effective quantum numbers $\nu_{N^+,v^{+}}$ associated to the vibrational thresholds of the closed ionisation channels.

\item {\it Computation of the cross-sections:}\\
For a given target cation on ro-vibrational level $N^+_i,v_i^+$ and for a given energy of the incident electron $\varepsilon$, the dissociative recombination and the ro-vibrational 
transition - elastic scattering, excitation, de-excitation -
the cross sections are computed using, respectively:
\begin{equation}\label{eqDR}
\sigma _{diss \leftarrow N_{i}^{+}v_{i}^{+}}^{N}=\frac{\pi }{4\varepsilon
}\frac{2N+1}{2N_{i}^{+}+1}\rho\sum_{l,\Lambda,d_{j}}
|S^{{N\Lambda}}_{d_{j},l N_{i}^{+}v^{+}_{i}}|^{2}
,
\end{equation}
\begin{equation}\label{eqVe}\begin{split}
&\sigma^N_{N_{f}^{+}v_{f}^{+} \leftarrow 
N_{i}^{+}v_{i}^{+}}=\frac{\pi }{4\varepsilon
}\frac{2N+1}{2N_{i}^{+}+1}\rho\\
&\times\sum_{l,l',\Lambda}
\left\vert
S^{N\Lambda}_{N_{f}^{+}v_{f}^{+}l',N_{i}^{+}v_{i}^{+}l}-\delta_{N^+_fN^+_i}\delta_{v^+_fv^+_i}\delta_{l'l}\right\vert
^{2}
,
\end{split}
\end{equation}
\noindent
where $\rho$ stands for the ratio between the spin multiplicities of the involved electronic states of CX and that of the target, CX$^+$.
\end{enumerate}

\vspace{.5cm}
\noindent
{\it
{b)  The account of {\em vibrational} structure and interactions, {\em neglecting the rotational ones}, for the target ion's {\em ground} electronic state and for the neutral's relevant electronic states~\cite{mezei2015}:
}
} 
 The target and the neutral systems are considered rotationally-relaxed, and $N^+$ is {not} a parameter in the steps 1-6 of case (a). In particular, the structure of the interaction matrix \Vmat introduced in step 1 is, in block form:
\begin{equation}\label{Vmat_1core}
\boldsymbol{\mathcal{V}} = \left( \begin{array}{cc}
\ 0 & \boldsymbol{\mathcal{V}}_{{\bar d}\bar v_{c}} \\
\boldsymbol{\mathcal{V}}_{\bar v_{c} \bar d} & \ 0 \\
\end{array} \right),
\end{equation}
 
 As for the step 7, the formulas (\ref{eqDR}) and (\ref{eqVe}) 
 become:
\begin{equation}\label{drxsec1}
	\sigma_{diss \leftarrow {v^+_{i}}} = \frac{\pi}{4\varepsilon} \sum_{{l,\Lambda}}
	\rho^{{\Lambda}} \sum_{j} \left|S^{{\Lambda}}_{d_j, {l}v^+_{i}}\right|^2.
\end{equation}
\begin{equation}\label{vexsec1}
	\sigma_{{v^+_{f}} \leftarrow  {v^+_{i}}} = \frac{\pi}{4\varepsilon}\sum_{{l,l'\Lambda}}
\rho^{{\Lambda}} \left|S^{{\Lambda}}_{ {l'}{v^+_{f}}, {l}{v^+_{i}} } - 
{\delta_{l'l}} \delta_{{v^+_{f}}{v^+_{i}}}\right|^2.
\end{equation}
\noindent
Here $\rho^{{\Lambda}}$ is the ratio between the spin and angular momentum multiplicities of the neutral and the target ion.

\vspace{.5cm}
\noindent
{\it
{c)  The account of {\em vibrational} structure and interactions, {\em neglecting the rotational} ones, for the target ion's  {\em ground} and {\em excited bound} electronic states,  and for the neutral's relevant electronic states~\cite{chakrabarti2018}:
}
} 

The structure of the interaction matrix \Vmat in block form, more complex than that in case (b) Eq. (\ref{Vmat_1core}) is, e.g., for the case of one ground bound electronic core 
and two excited bound electronic cores 
the following:
\begin{equation}\label{Vmat2}
\boldsymbol{\mathcal{V}} = \left( \begin{array}{cccc}
\ 0 & \boldsymbol{\mathcal{V}}_{{\bar d}\bar v_{c_1}} & \boldsymbol{\mathcal{V}}_{\bar d \bar v_{c_2}}
& \boldsymbol{\mathcal{V}}_{\bar d \bar v_{c_3}}\\
\boldsymbol{\mathcal{V}}_{\bar v_{c_1} \bar d} & \ 0 & \boldsymbol{\mathcal{V}}_{\bar v_{c_1} \bar v_{c_2}}
& \boldsymbol{\mathcal{V}}_{\bar v_{c_1} \bar v_{c_3}}\\
\boldsymbol{\mathcal{V}}_{\bar v_{c_2} \bar d} & \boldsymbol{\mathcal{V}}_{\bar v_{c_2} \bar v_{c_1}} & \ 0
& \boldsymbol{\mathcal{V}}_{\bar v_{c_2} \bar v_{c_3}}\\
\boldsymbol{\mathcal{V}}_{\bar v_{c_3} \bar d} & \boldsymbol{\mathcal{V}}_{\bar v_{c_3} \bar v_{c_1}} & \boldsymbol{\mathcal{V}}_{\bar v_{c_3} \bar v_{c_2}}
& \ 0\\
\end{array} \right),
\end{equation}
\noindent where the collective indices $\bar d$, $\bar v_{c_i}$, $i=1,2,3$ span the ensembles of all individual indices connected to the dissociation channels and ionisation channels built on $c_1$ (ground), $c_2$ and $c_3$ (excited) ion cores~\citep{annick1980,carata2000,novotny2009,wafeu2011,
motapon2014,mezei2015,faure2017,moulane2018,chakrabarti2018}. 

Here, besides the {\it Rydberg-valence} interactions, represented by $\boldsymbol{\mathcal{V}}_{{\bar d}\bar v^+_{c_1}} $,  
$\boldsymbol{\mathcal{V}}_{{\bar d}\bar v^+_{c_2}} $ and  $\boldsymbol{\mathcal{V}}_{{\bar d}\bar v^+_{c_3}} $,  like in (\ref{Vmat_1core_rot}) and (\ref{Vmat_1core}),
one may notice the appearance of the {\it Rydberg-Rydberg} ones, represented by 
$\boldsymbol{\mathcal{V}}_{\bar v^+_{c_1} \bar v^+_{c_2}}$, $\boldsymbol{\mathcal{V}}_{\bar v^+_{c_1} \bar v^+_{c_3}}$ and $\boldsymbol{\mathcal{V}}_{\bar v^+_{c_2} \bar v^+_{c_3}}$.

\section{{Application to carbon-based diatomic molecular systems}}

The molecular data necessary to model the dissociative recombination and the ro-vibrational (de)excitation {given in the diabatic representation} are:

(i)  the potential energy curve (PEC) of the ground state of the ion, CX$^+$,

(ii)  the  PECs of the excited attractive states energetically close to the ion's ground state one,  CX$^{+*}$,

(iii) the PECs of the valence dissociative states of the neutral CX$^{**}$ interacting with the ionization continua, 

(iv) the PECs of the Rydberg bound states CX$^{*}$ associated to the ionization continua and situated below the 
ion states CX$^+$ or/and CX$^{+*}$, which can be conveniently described by smooth $R$-dependent quantum defects, predissociated by the CX$^{**}$ states and being furthermore subject to interseries Rydberg-Rydberg interactions.

(v) the electronic couplings between the valence dissociative states and the Rydberg manifolds, as well as the Rydberg-Rydberg electronic couplings whenever multiple Rydberg manifolds are present. 

\begin{figure}[t]
     \centering
     \includegraphics[width=0.75\linewidth]{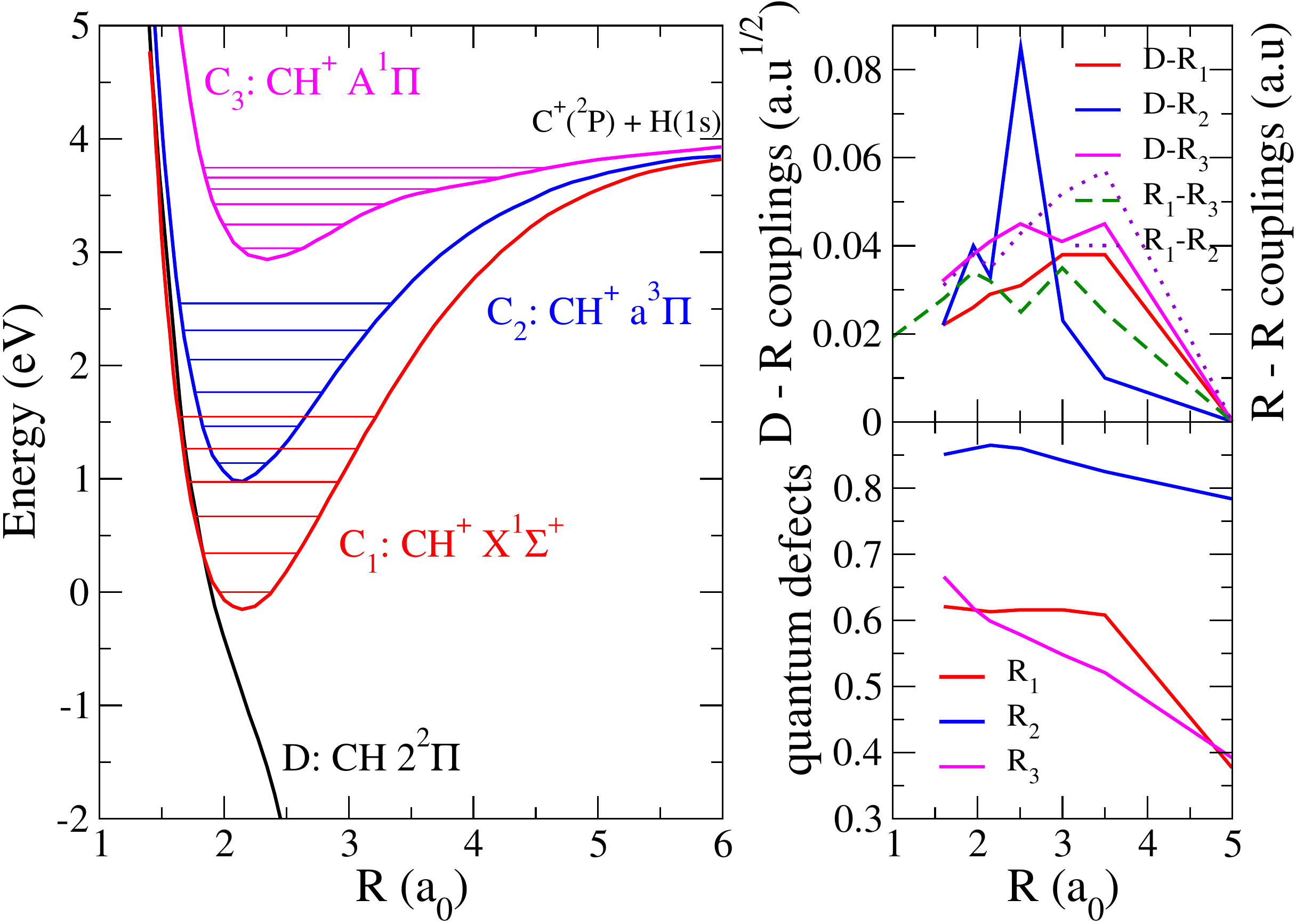}
 \caption{{Molecular data sets relevant for the dissociative recombination of CH$^+$ compiled from Figs.~$2-5$ of 
 Ref.~\citep{carata2000}. Left panel: the red, blue and magenta continuous lines stand for the ground (C$_1$: X $^1\Sigma^+$) and the lowest two excited (C$_2$: a $^3\Pi$ and C$_3$: A $^1\Pi$) electronic states of the ion, having the C$^+(^2$P) $+$ H($1s$) dissociation limit, whereas the black continuous (D) line gives the dissociative autoionizing state (2 $^2\Pi$) of the neutral. Upper right panel: electronic couplings. Red, blue, and magenta lines between the valence state (D) and the states belonging to the Rydberg series R$_1$, R$_2$ and R$_3$ built on the C$_1$, C$_2$ and C$_3$ ion cores respectively. The violet dotted and  the green dashed lines stand for the couplings of the R$_1$ Rydberg series to the R$_2$ one and that of the R$_1$ Rydberg series to the R$_3$ one respectively. The coupling between R$_2$ and R$_3$ were considered zero. Lower right panel: quantum defects for the Rydberg series based on the ground and excited ion cores.
}
}
  \label{fig:CH0}
\end{figure}

\noindent A representative example of molecular data needed for the MQDT calculations - relevant for the DR of CH$^+$ - is shown in Fig.~\ref{fig:CH0}.

The DR cross section is extremely sensitive to the position of the neutral dissociative states with respect to that of the target ion. 

Several methods are available to provide all the necessary molecular data with the desired accuracy. Among these are R-matrix theory~\citep{tennyson2010}, the complex Kohn variational method~\citep{rescigno1995}, the quantum defect methods~\citep{jungen1996}, the interpretation of of spectroscopic data~\citep{greene1985} and the block diagonalization method~\citep{pacher1988,kashinski2017}.

Following the molecular data preparations as described above we have performed a series of MQDT calculations of cross section for the DR and its competitive processes involving relevant cations, as follows.

\subsubsection{CH$^+$}

CH$^+$ was first found in the interstellar molecular clouds in $1941$ by Douglas and Herzberg~\citep{douglas1941}. Since then, its absorption lines have been observed towards many background stars, demonstrating the omnipresence of this simple carbon hydride in the diffuse interstellar medium (ISM). {This outstanding abundance is so far unexplained. }

The most probable formation channel of the CH$^+$ cation is the hydrogen abstraction reaction $\mbox{C}^++\mbox{H}_2 \rightarrow \mbox{CH}^++\mbox{H}$, which is endothermic. In order to be formed in a sufficient abundance for observation, alternative energy sources were suggested such as turbulent dissipation, shocks, or shears (see e.g. Valdivia {\it et al.}~\citep{valdivia2017}, and references therein). Furthermore, it was found that, in photon dominated regions, the ro-vibrationally excited H$_2$ reservoir can provide an alternative route to overcome the endothermicity of the formation reaction~\citep{agundez2010}, since the rotational and vibrational energies are as effective as the translational energy in promoting this type of reaction~\citep{zanchet2013}.

This reaction is in competition with the low-density process known as radiative association~\citep{herbst2008}, $\mbox{C}^++\mbox{H} \rightarrow {\mbox{CH}^+}^* \rightarrow \mbox{CH}^++h\nu$ in which the collision complex is stabilised by emission of a photon.

Yet another ion-molecule reaction able to fix atomic carbon into a molecular form and thus start the build-up of organic molecules, according to \citep{herbst2008,savin2015} is the proton-exchange reaction: $\mbox{H}_3^+ + \mbox{C} \rightarrow \mbox{CH}^++\mbox{H}_2$. 

It is obvious from the previous examples that the detailed knowledge of the CH$^+$ molecular cation chemistry can provide unique physical insights into the modelling of the different ISM environments. CH$^+$ is one of the major building block of the hydrocarbon chemistry~\citep{mccallphd}, that leads through radiative association to more complex radicals like CH$_3^+$ or CH$^+_5$, key pieces of the carbon chemistry of ISM that leads to COMs (for a representative sample see eqs.~(\ref{com1})-(\ref{com4})). Thus the full understanding of both production and loss mechanisms, as well as the competition between the radiative processes, the destruction of the ion and the collisional excitation processes is needed to be known in detail.

CH$^+$ is easily destroyed by reactions with electrons and hydrogen atoms, and also by reactions with H$_2$ molecules. Here we {show and discuss} cross sections on the electron induced destruction pathways and competitive processes like ro-vibrational excitations.

The first complete potential energy curves of CH$^+$ and CH relevant for DR of CH$^+$ were produced by Giusti-Suzor and Lefebvre-Brion~\citep{giusti1977}, where the authors did not find a favorable crossing of the ion curve with a neutral one. This finding led to a slow DR rate coefficient. Similar results were obtained by the earlier theoretical calculations of Bardsley and Junker~\citep{bardsley1973}.
The most complete calculation was performed by Takagi~{\it et al}~\citep{takagi1991}, 
who obtained the relevant molecular structure data using configuration mixing methods, and provided reliable DR cross sections and low temperature rate coefficients using a different version of MQDT in reasonable agreement with that predicted by~\citep{bardsley1973} and the experimental values given by Mitchell and McGowan~\citep{mitchell1978}. 

On the experimental side, this first merged-beam measurement of DR for CH$^+$~\citep{mitchell1978} was followed by a new revised one of the same team~\citep{mul1981}, resulting in slightly larger values. All these result were later recompiled in~\citep{mitchell1990}. The most detailed experimental study of CH$^+$ DR was performed on a heavy-ion storage-ring equipment, resulting in cross sections, product branching ratios and angular distributions, was given by Amitay {\it et al}~\citep{amitay1996}. Their measurements showed several broad and prominent resonances that were tentatively attributed to the capture of the incident electron into core-excited Rydberg states. In order to understand and characterise the broad resonances, Carata {\it et al}~\citep{carata2000} performed, on a set of new molecular structure data of CH$^+$ and CH, new calculation by including in the available MQDT-DR approach the effect of these core-excited states in a first-order perturbative approximation. 

\begin{figure}[t]
     \centering
     \includegraphics[width=0.75\linewidth]{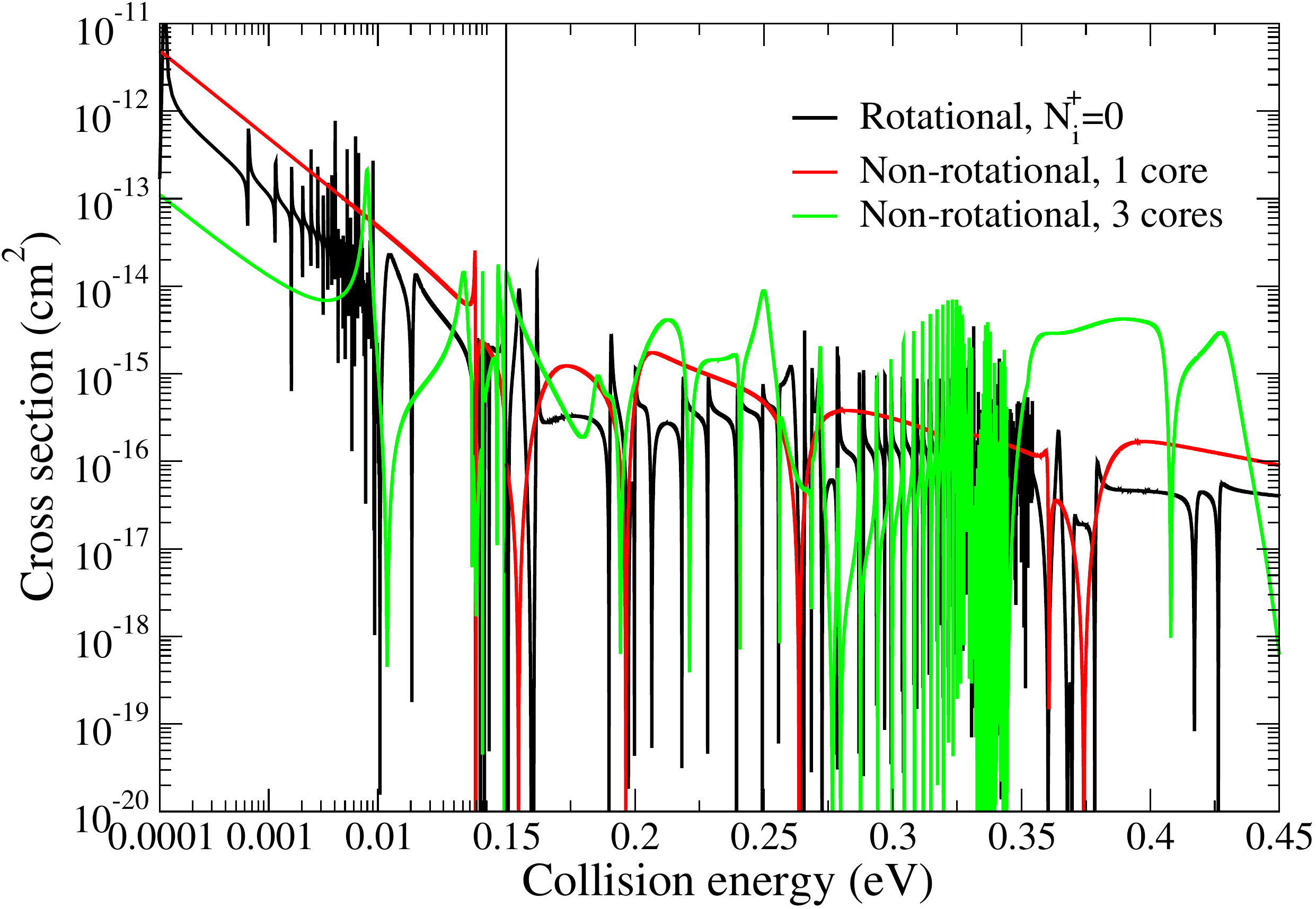}
 \caption{Cross sections for the dissociative recombination of the vibrationally relaxed CH$^+$ by electrons as function of the collision energy.  The initial-state-specific ($N_i^+=0$) including rotation are compared to two non-rotational calculations using the ground (red curve) and all three ionic cores (green curve). In the left panel the results are shown in log-log scale, while in the right one in log-linear scale. 
}
  \label{fig:CH1}
\end{figure}

Here we discuss two different sets of MQDT calculations~\citep{faure2017,chakrabarti2018, mezei2019} which have been performed for this molecular cation making use of the same molecular data sets, {shown in figure~\ref{fig:CH0} and} presented originally in Ref.~\citep{carata2000}.

In a first step, relying on the ground electronic state of the ion only, and valid for very low energy, the dissociation state of the neutral and the interaction between the ionization and the dissociation continua, we have computed the interaction-, reaction- and scattering matrices, and produced the DR cross sections for the $11$ lowest rotational levels of CH$^+$ in its ground electronic and vibrational state, using the version ($a$) of our MQDT method~\citep{mezei2019}.

In the second step of our MQDT calculations, devoted to higher incident electron energy, we focused on the importance of the excited bound ionic cores, using the approach  ($c$), {adopting a second-order perturbative solution for eq.~(\ref{eq:solveK}) in contrast to the first-order treatment employed in ref.~\citep{carata2000}.}
Indeed, CH$^+$ has several bound excited states whose ionisation continua are coupled to the ionisation continuum of the ground core and to the neutral dissociative states. For the energy range characterising the incident electron in the present work, two such excited states are relevant, i.e. those of $a\,^{3}\Pi$ and $A\,^{1}\Pi$ symmetry, which we respectively call core $2$ and core $3$. The neutral $2\, ^{2}\Pi$  dissociative state is coupled to the ionisation channels of the three ion cores and is mainly responsible for driving the low energy DR mechanism.

We have used in total  $42$ ionisation channels associated with $19$ vibrational levels of the CH$^+$ ($X ^{1}\Sigma^{+}$) ground state ($c_1$) and $14$ and $9$ vibrational levels of CH$^+$ ($A ^{3}\Pi$) ($c_2$) and CH$^+$ ($a ^{1}\Pi$) ($c_3$) respectively. The incident electron energy range is $0.01-0.5$ eV, which is typical for the interstellar environments. The $l = 1$ (p) partial wave was considered for the incident/Rydberg electron.

The rate coefficients were determined at kinetic temperatures between $10$ and $3000$ K.

Figure~\ref{fig:CH1} shows the DR cross section for $1$ core$/1$ dissociative state including rotation (black curve), 
$1$ core$/1$ dissociative state excluding rotation ({MQDT approach $b$}, red curve) and $3$ core$/1$ dissociative state 
excluding rotation ({MQDT approach $c$}, green curve) respectively. 

The effects of inclusion of rotation (black vs. red curves) and of the excited cores (red vs. green curves) is striking. New Fano type-resonances appear in the cross section due to the ro-vibrational levels of the super-excited Rydberg states of the neutral, correlating to the ground and excited ion states ({\it indirect} mechanism). The quantum interference between the {\it direct} and {\it indirect} mechanisms lead to shifted mean cross section values.  

\begin{figure}[t]
     \centering
      \includegraphics[width=0.75\linewidth]{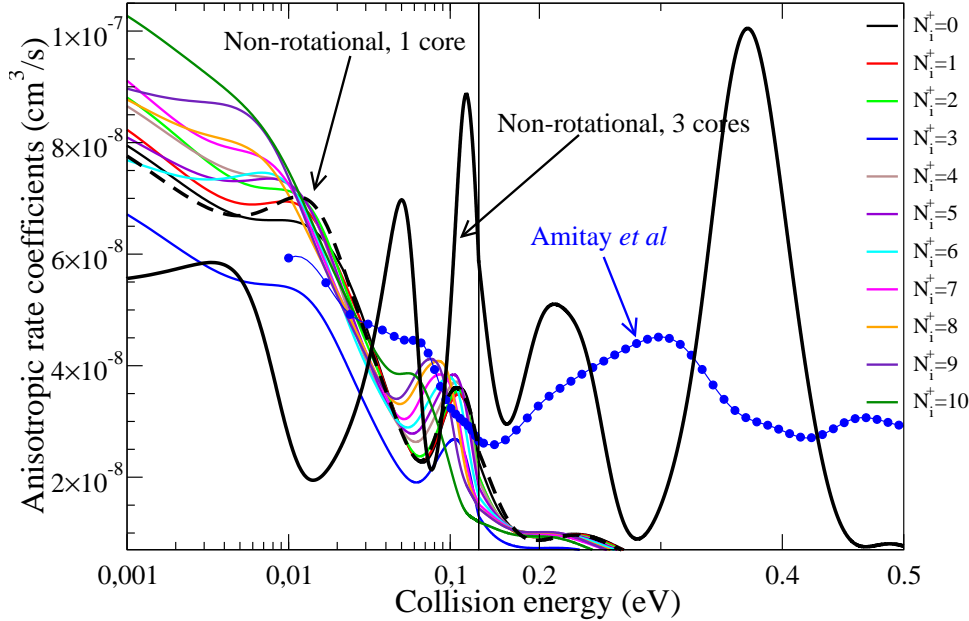}
 \caption{Anisotropic rate coefficients for the dissociative recombination of the vibrationally relaxed CH$^+(N_i^+)$ as function of the kinetic energy of the electrons. The initial-state-specific rate coefficients {(coloured continuous lines)} are compared to two non-rotational calculations using the ground {(black dashed line)} and all three ionic cores { (thick black line)} and to the experimental results of Amitay {\it et al.}~\citep{amitay1996} {(blue line with circles)}. {The anisotropic Maxwell averaging was done according eq. (1) of ref.~\citep{schippers2004} using $T_\perp=17$ meV and $T_\parallel=0.5$ meV electron temperatures}. In the left panel the results are shown in log-log scale, while in the right one in log-linear scale.
 }
  \label{fig:CH2}
\end{figure}

The best comparisons between the MQDT calculations and the storage-ring experiments can be made starting from the convoluted cross sections shown in figure~\ref{fig:CH2}. Here we have used anisotropic Maxwellian distribution taking into account the experimental conditions of Amitay {et al.}~\citep{amitay1996}. Whereas the rotational effects are relevant for low collision energies, the excited core effects become important for  higher energy, where the { strong}  resonances revealed by the experiment are qualitatively well reproduced by our calculations. { Meanwhile, the positions and the intensities of the experimentally observed resonance profiles are not yet quantitatively well reproduced. We hope to remove this disagreement by using in the near future more accurate molecular structure data on the super-excited electronic states of CH, and by taking into account simultaneously the effects of core-excited resonances and of the rotational effects.}

\begin{figure}[t]
     \centering
     \includegraphics[width=0.75\linewidth]{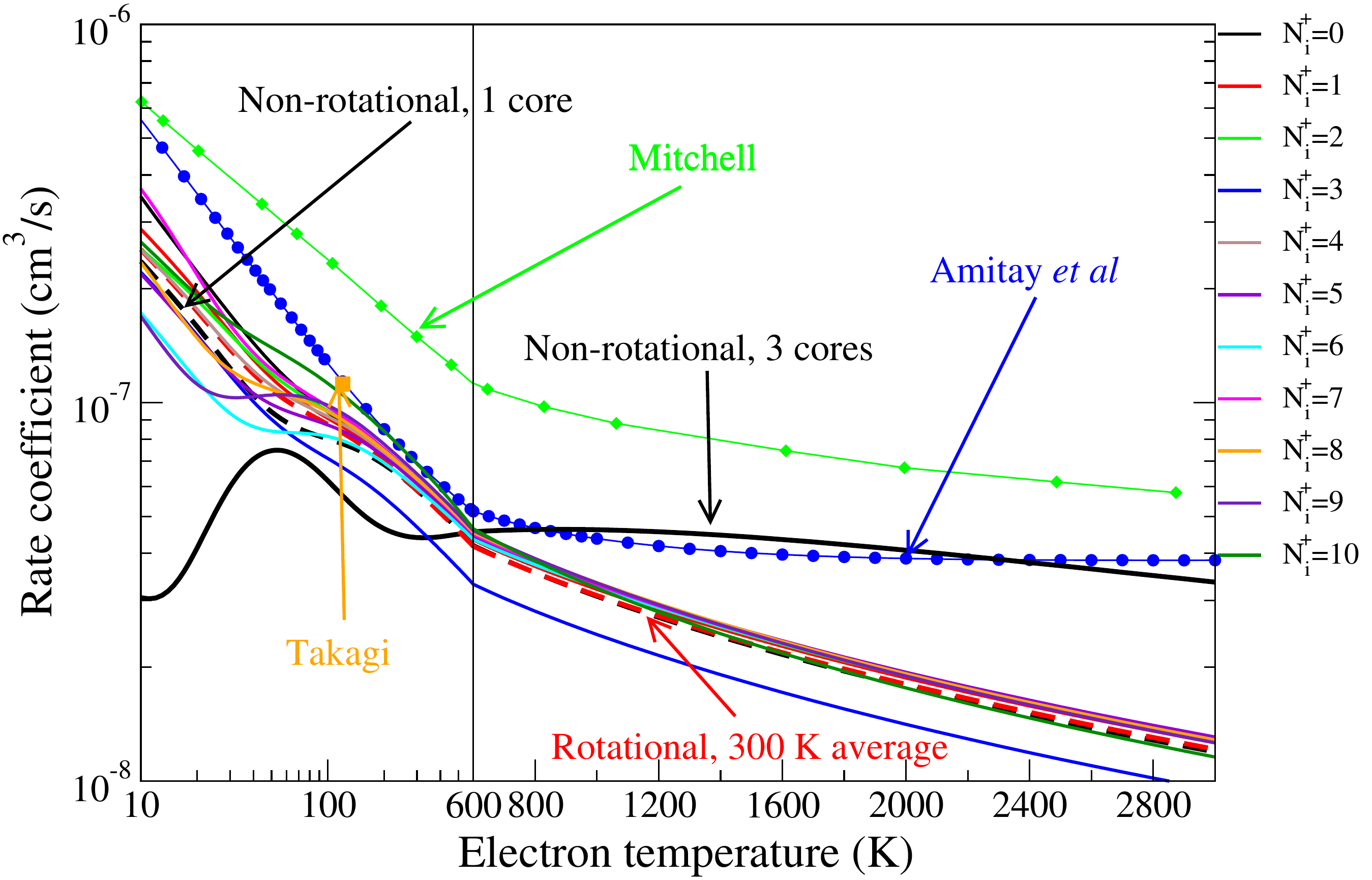}
 \caption{Maxwellian rate coefficients for the dissociative recombination of the vibrationally relaxed CH$^+(N_i^+)$ by electrons as function of the kinetic temperature. The initial-state-specific {(coloured continuous lines)} and $300$ K thermal rate coefficients {(red dashed line)} from this work are compared to two non-rotational calculations using the ground {(black dashed line)} and all three ionic cores {(thick black line)} and to the experimental results of Amitay {\it et al.}~\citep{amitay1996} {(blue line with circles)} and Mitchell~\citep{mitchell1990} {(green line with diamonds)}. {The rate constant calculated at $T=120$ K electron temperature by Takagi {\it et al}~\citep{takagi1991} with a different MQDT approach is given in orange square.} 
}
  \label{fig:CH3}
\end{figure}

And finally in Fig.~\ref{fig:CH3} (using the same colour-code as Fig.~\ref{fig:CH2}), the initial-state-specific {(coloured lines)}
DR rate coefficients are plotted as functions of the kinetic temperature. These rate coefficients were obtained by averaging the cross sections over isotropic Maxwell-Boltzmann velocity distributions. The thermal average at a rotational temperature of $300$ K is also shown {(red dashed line)} and it is compared to two sets of non-rotational MQDT calculations using the ground {(black dashed line)} and all three excited cores {thick black line} and to two sets of experimental rate coefficients. The first was obtained from a thermal average of the DR cross sections measured by Amitay et al.~\citep{amitay1996}
(blue line with circles), where the ions were assumed to be thermalized at the ambient temperature of the storage ring ($300$ K). In order to be able to compare the calculated rate coefficients with the measured ones at low electron temperatures, we have extrapolated the experimental cross sections towards low collision energies assuming a Winger-type behaviour: $\sigma(\epsilon)=a_0/\epsilon$, where the fitting parameter were taken: $a_0=1.559\times10^{-16}$ cm$^2\cdot$eV. The second is the experimental recommendation of Mitchell~\citep{mitchell1990} {(green line with diamonds)}. 

The theoretical thermal average at $300$ K is significantly lower but the agreement with Amitay {\it et al.}~\citep{amitay1996} is within a factor of $2$ up to $\sim1000$ K. One should notice that the variation of the DR rate coefficients is not monotonic with $N_i^+$, i.e. there is no particular trend with increasing $N_i^+$. 
Their dependence on the rotational excitation of the target ion is
weak, and all the initial-state-specific rates agree to within a factor of $2$. Here, the DR Maxwell rate coefficients computed for the lowest $11$ rotational levels of the ground vibrational level, are  compared with two non-rotational cases and the averaged rate $300$ K. Fig.~\ref{fig:CH3} shows stronger dependence on the initial rotational levels of the molecular cation for temperatures up to $200$ K. In this region one can observe the largest differences between the averaged and non-rotational with ground core only rates. The second set of non-rotational rates (three ionic cores) are very convincing on the minor importance on the {rotational} effects for this particular molecular cation. Above this temperature, the rotational effect  {is} even less pronounced, the non-rotational rate coefficient (one core only) and that one of averaged at $300$ K are {on} top of each other. This is partially due to the peculiar behaviour of the rate coefficient belonging to the $N_i^+ = 3$ rotational state. 

The best agreement  between the { non-rotational} MQDT calculations { for three cores} and the { thermally convoluted} experimental results of Amitay {\it et al}~\citep{amitay1996} is achieved for temperatures higher than $600$ K, { although the comparison with the anisotropic Maxwell rate coefficients shows only qualitative agreement} . This very good agreement lasts up to $2800$ K, and frames the region where the excited core effects are really important. 

\begin{figure}[t]
     \centering
     \includegraphics[width=0.75\linewidth]{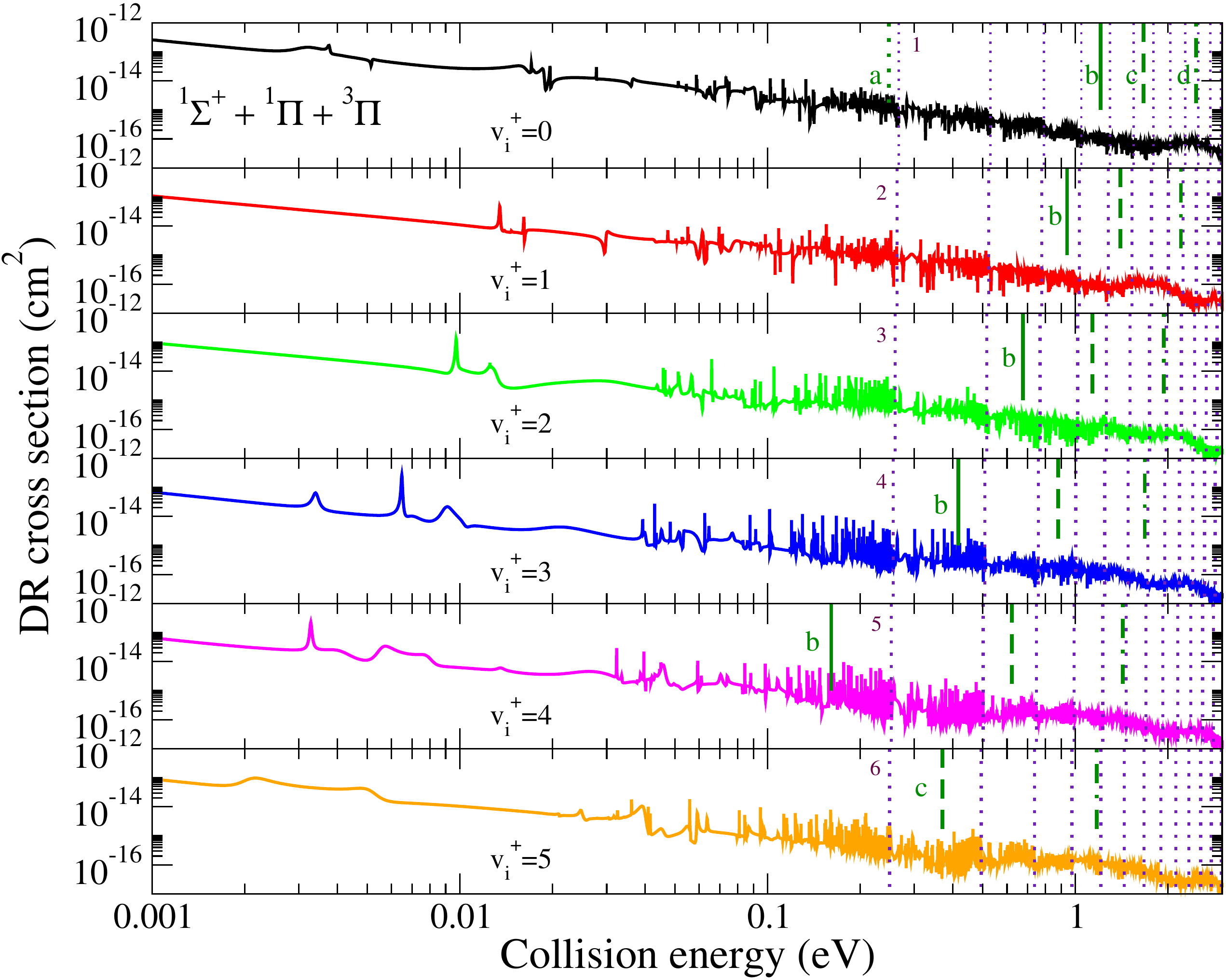}
 \caption{Dissociative Recombination  of CO$^+$ on its lowest six vibrational levels ($v_i^+$=0, 1, 2, 3, 4 and 5):  cross sections summed-up over all the relevant symmetries - see ~\citep{mezei2015}. The dotted vertical indigo lines are the different ionisation thresholds, given by the vibrational levels of the molecular ion. The first ionisation thresholds are indicated on the figures. The dark-green shorter vertical lines stand for the different dissociation limits measured from the initial vibrational levels of the ion, as follows: the dotted line (a) stand for the C($^1$D)$+$O($^1$D) limit, the solid lines (b) for the C($^3$P)$+$O($^1$S) one, the dashed lines (c) for C($^1$S)$+$O($^1$D) , and the dashed-dotted lines (d) for  C($^1$D)$+$O($^1$S).
}
  \label{fig:CO1}
\end{figure}

\subsubsection{CO$^+$}

The carbon monoxide ion CO$^+$ is one of the most abundant ions detected in the interstellar medium~\citep{fuente1997}, in the coma and in the tail region of comets, and  is of key relevance for the Martian atmosphere \citep{fox1999}. It has been detected by several spacecraft missions to different comets, and is thought to be formed by photoionization and electron impact ionization of CO and CO$_2$ molecules~\citep{vojnovic2013}. 

The high density of electrons and molecular ions in the cometary coma and some parts of ISM facilitates dissociative recombination. Moreover,  this process plays an important role in producing numerous carbon and oxygen atoms in metastable states. 

The first experimental results on the dissociative recombination of carbon monoxide was obtained in the late sixties by Mentzoni and Donohoe~\citep{mentzoni1968} in a dc discharge afterglow measurement. This was followed by the merged beam experiment of Mitchell and Hus~\citep{mitchell1985} in 1985 and by the flowing afterglow experiment of Geoghegan {\it et al}~\citep{geoghegan1991} in 1991. The most recent and complete experimental results were obtained by Rosen {\it et al}~\citep{rosen1998} on the CRYRING storage ring equipment.
The first and pioneering theoretical study providing rate coefficients was done by Guberman~\citep{guberman2007} in 2007 that was followed in 2013 by an updated quantum chemistry calculation~\citep{guberman2013} providing the most relevant molecular states for DR of CO$^+$. 

By using the R-matrix based calculations of { Chakrabarti and Tennyson}~\citep{chakrabarti2006, chakrabarti2007} completed by other {\it ab initio} quantum chemistry calculations~\citep{vazquez2009,lefebvre2010} we have managed to set up a molecular data set containing the three most important symmetries contributing to the DR and its competitive processes, namely the $^1\Sigma^+$, $^1\Pi$ and $^3\Pi$ ones, and considering four dissociative states for each symmetry. {The calculations were performed using the MQDT version $b$.} For each available dissociative channel, we have considered its interaction with the most relevant series of Rydberg states, that is, s, p, d and f, for the $^1\Sigma^+$, symmetry, and s, p and d for the $^1\Pi$ and $^3\Pi$ symmetries.

The DR cross sections are displayed in Fig.~\ref{fig:CO1}~\citep{mezei2015,moulane2018}. They are characterized by resonance structures due to the temporary captures into vibrational levels of Rydberg states embedded in the ionization continuum (closed channels, indirect process), superimposed on a smooth background originating in the direct process.

The ionization thresholds (vibrational levels of the molecular ion) shown as dotted vertical lines in Fig.~\ref{fig:CO1} act as accumulation points for these Rydberg resonances. Moreover, the asymptotic limits of the dissociation channels opening progressively are shown with shorter dark-green vertical lines, corresponding to the atomic pairs of states C($^1$D) + O($^1$D), C($^3$P) + O($^1$S), C($^1$S) + O($^1$D), and C($^1$D) + O($^1$S). We notice that the C($^3$P) + O($^3$P) and C($^1$D) + O($^3$P) limits are open at zero collision energy.

\begin{figure}[t]
     \centering
     \includegraphics[width=0.75\linewidth]{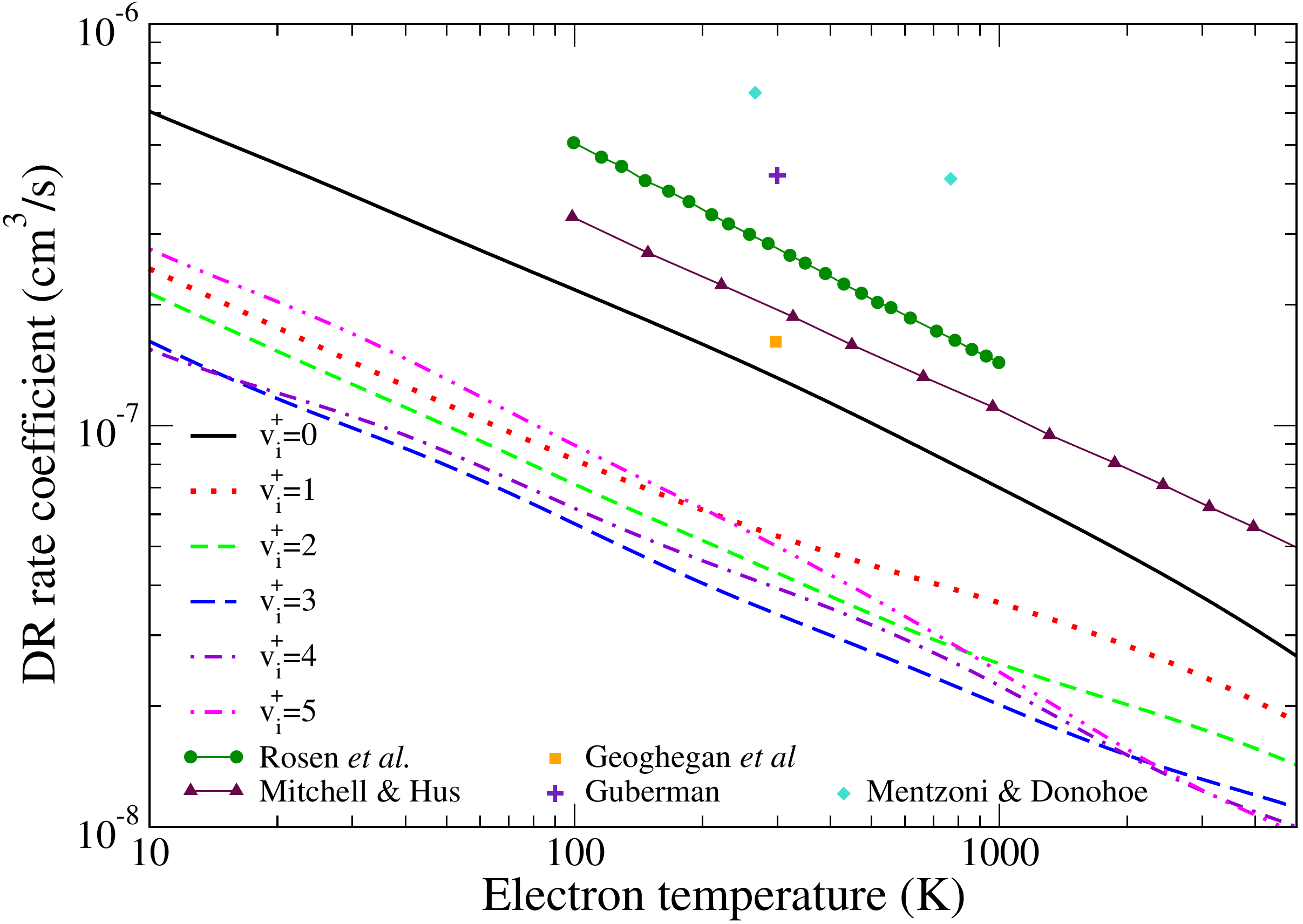}
 \caption{Dissociative Recombination  of CO$^+$ on its lowest six vibrational levels ($v_i^+$=0, 1, 2, 3, 4 and 5):  Maxwell rate coefficients summed-up over all the relevant symmetries, coloured lines with different line styles - see ~\citep{mezei2015}.  The green line with full circles are the experimental data measured on CRYRING, Stockholm~\citep{rosen1998}, {the maroon line with triangles are the estimates provided by Mitchell and Hus~\citep{mitchell1985}, the turquoise diamonds are the measurements of Mentzoni and Donohoe~\citep{mentzoni1968}, the orange square is the experimental rate of Geoghegan {\it et al}~\citep{geoghegan1991}, and finally the indigo cross is the theoretical estimate of Guberman~\citep{guberman2007}.}
}
  \label{fig:CO2}
\end{figure}

The DR of  CO$^+$ vibrationally relaxed  (top panel of Figure~\ref{fig:CO1}) is by far dominant: it is about four times larger below $700$ meV and above $2$ eV, while in-between, the maximum deviation among all the cross-sections is smaller than a factor of two. At low energy, one can observe a systematic decrease of the total cross-section, except for $v^+_i = 5$ (bottom panel of the same figure). In this latter case, the PECs of the two open valance states of $^1\Sigma^+$ symmetry have favorable crossings with the ion PEC - symmetry with the largest valence-Rydberg electronic couplings (see Fig. 2 from~\citep{mezei2015}) - leading to an increase of the cross section.

Another interesting feature is the revival of the cross section at high energy, due to the opening of dissociation states.

In spite of the overall factor $2$ between { our MQDT calculated rate coefficients} and {all experimental results (except the higher placed rates of Mentzoni and Donohoe due to the presence of CO-clusters)}, the {\it shape} agreement achieved over a significant range of energies and temperatures, illustrated in Fig.~\ref{fig:CO2}, but also, in more detail, in Figs. 4-6 from~\citep{mezei2015}, is satisfactory for this diatomic system with many electrons.

\subsubsection{CF$^+$}

Whereas the DR of fluorine-containing molecular cations controls the ionization state and the chemical composition
of many plasmas used in electronic processing~\citep{denpoh2000, georgieva2003, torregrosa2004, mezei2016, slava2018}, it also does, especially in the case of CF$^+$, for the  fluorine chemistry in the cold interstellar medium~\citep{neufeld2006,roueff2009}. 

{ Very little has been done regarding the quantum chemistry of the super-excited states of CF and the DR CF$^+$.} The relevant molecular PECs and electronic couplings shown in Figures 4. and 5. in Ref.~\citep{novotny2009} have been calculated by the complex Kohn variational method \citep{rescigno1995}. The molecular data set contained one ion core and one dissociative molecular states for each of the three $^2\Sigma$, $^2\Pi$ and $^2\Delta$ symmetries. The MQDT calculations - {using approach $b$}~\citep{novotny2009} - were performed without including the Rydberg bound states responsible for the {\it indirect} process. The results were compared with measurements performed on the Test Storage Ring (TSR), Heidelberg.

\begin{figure}[t]{ 
     \centering
     \includegraphics[width=0.75\linewidth]{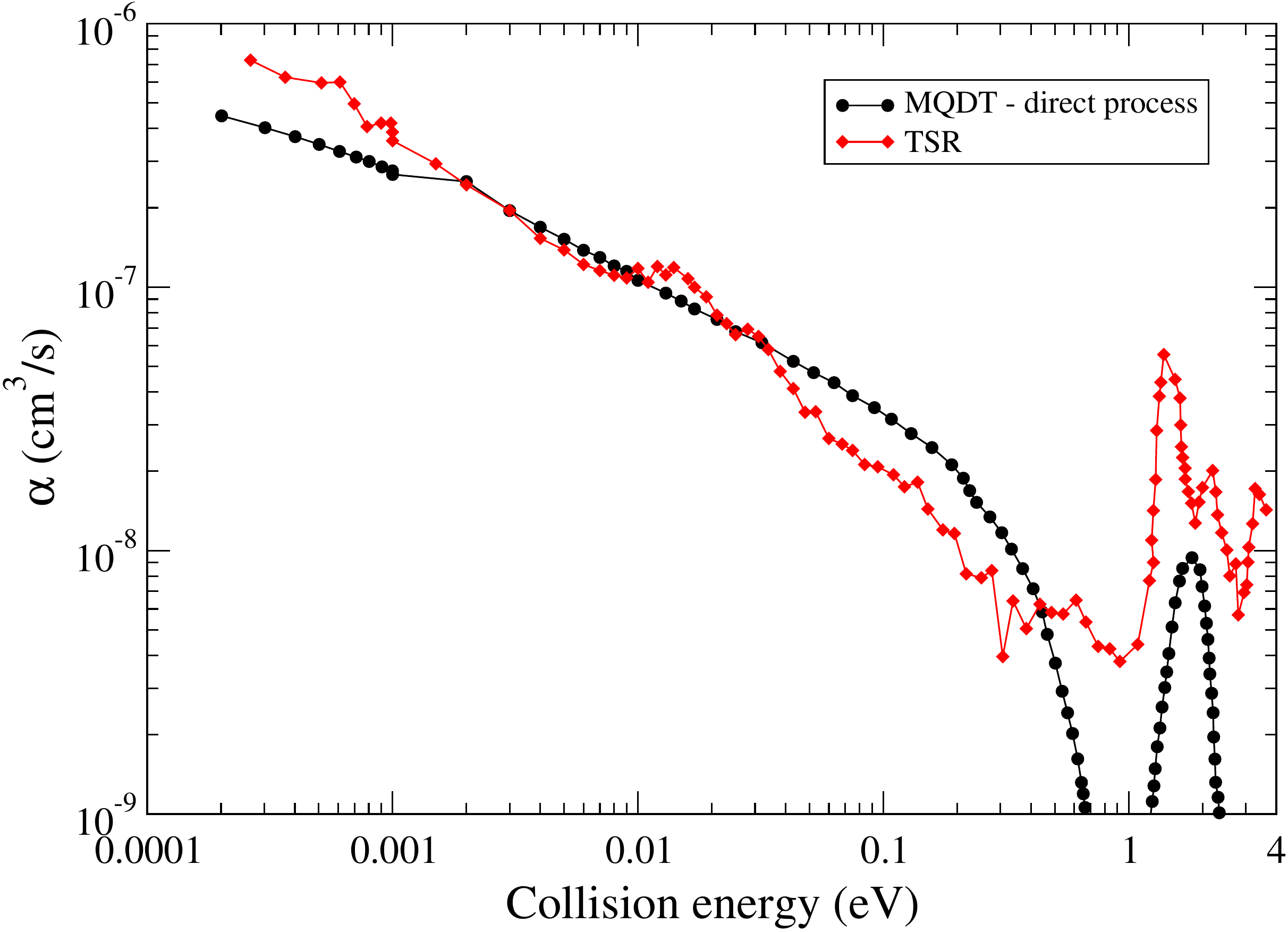}
 \caption{DR anisotropic rate coefficients for the vibrationally relaxed ($v_i^+=0$) CF$^+$ molecular ion compared to TSR experiment~\citep{novotny2009}.}
\label{fig:CF1}
 }
\end{figure}

\begin{figure}[b]{ 
     \centering
     \includegraphics[width=0.75\linewidth]{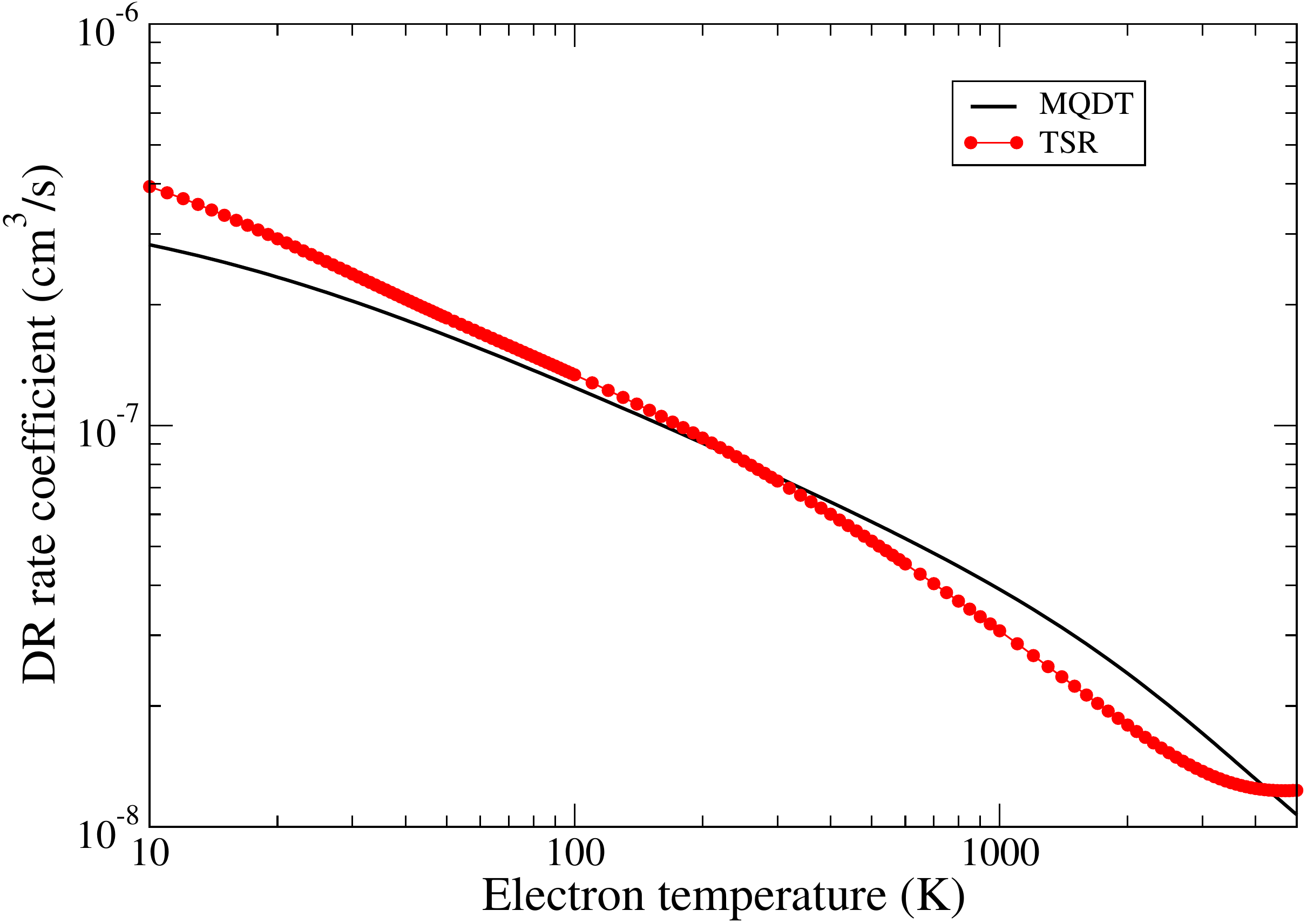}
 \caption{DR Maxwell isotropic rate coefficients for the vibrationally relaxed ($v_i^+=0$) CF$^+$ molecular ion compared to TSR experiment~\citep{novotny2009}.}
\label{fig:CF2}
 }
\end{figure}

The anisotropic Maxwell rate coefficients for both experiments and theory are reported on Figure~\ref{fig:CF1}. The agreement is very good up to an electron energy close to $1$ eV. At higher energies the discrepancies observed can be accounted for the {indirect mechanism}, for the higher dissociative {and/or excited ion} curves that have not been considered in the present treatment. The very good agreement is even more visible for the isotropic rate coefficients presented in Fig.~\ref{fig:CF2}. 
\clearpage

\clearpage
\subsection{Numerical details}
The key numerical parameter determining the memory used and the running time is the dimension of the matrices involved in computation:
the interaction matrix 
$\boldsymbol{\mathcal{V}}$, 
(eqs. (7), (13) and (16)), 
the reaction matrix $\boldsymbol{\mathcal{K}}$
(eq. (8)), 
the generalized scattering matrix 
$\boldsymbol{{X}}$
(eq. (9)) 
and 
the physical scattering matrix   $\boldsymbol{\mathcal{S}}$  
(eq. (10)).
The dimension of the matrices is given by the number of reaction channels - dissociation channels and ionization ones. For example, for the  DR of CO$^+$ (see above),  for a given electronic symmetry (from the three relevant ones) of the CO system, we considered 212 ionization channels, associated, each of them, to one of the 53 vibrational levels of the ion and to one of the 4 partial waves of the incident electron. 

An other key parameter is the dimension of the grid of energies of the incident electrons. Using the same example of the DR of CO$^+$, in order to take into account the indirect process characterized by fine Rydberg resonances, we had to use a step of $10^{-2}$ meV, which, for an incident electron energy up to .5 eV, implies 50,000 energy-points. 

These features correspond in average to runs on a "standard" desktop computer (Xeon CPU with 6 cores) taking between one and two weeks for each initial vibrational  level of the target ion. We expect improvement in the overall running time following parallelization of the numerical routines/methods over the energy-grid-points.

\section{Polyatomic molecules}

\subsection{Vibrational dynamics in DR of polyatomic molecules}

The multi-dimensional nature of vibrational motion of polyatomic ions makes the theoretical study of the dissociative recombination and rovibrational excitation of such ions much more difficult \cite{Kokoouline01,Kokoouline03a,florescu06,larsson08}
 compared to the case of diatomic ions discussed above. Once the electron is captured, the complexity of the dissociation processes increases not only with the multi-dimensional vibrational and dissociative dynamics, but also with the symmetry of the neutral system \cite{Kokoouline01,Kokoouline03a,douguet12b,douguet12a}. 
{In particular, this complexity applies to the spectacular role of the indirect process, already discussed and illustrated 
in the diatomic case, especially for the CH$^+$ DR.}  
For polyatomic ions with high symmetry, such as linear ions \cite{mikhailov06,douguet08b} or ions having three or more identical nuclei \cite{douguet12b,douguet12a}, the non-Born-Oppenheimer coupling between the incident electron and the rovibrational degrees of freedom could be non-trivial \cite{Kokoouline01,Kokoouline03a,Kokoouline03c,mikhailov06,douguet08b} and special attention should be given to the construction of the reaction and scattering matrices. However, the main ideas of the MQDT approach described above {for the diatomic ions} can be applied to polyatomic ions as long as the symmetry of {the} total system is accounted for in modeling the  non-Born-Oppenheimer coupling, which is responsible for VE and DR processes.

Over the recent years, three major approaches were developed to account for vibrational dynamics of the target polyatomic ion and the neutral molecule formed after the recombination step in the DR process. 

The simplest approach is based on the idea that for certain ions, only one vibrational degree of freedom is responsible for {the} dynamics \cite{orel2005wave,hickman2005dissociative}. Whenever this approximation is reasonable, one can apply almost all techniques developed for diatomic ions.

The second approach was initially developed for the DR and VE treatment of the H$_3^+$ ion and it uses the properties of the hyper-spherical coordinates \cite{Kokoouline01,Kokoouline03a,Kokoouline03c,kokoouline04a,samantha14}. This approach is very useful for systems with only a few (2-3) vibrational degrees of freedom that should be taken into account explicitly. The idea of the approach can be summarized as follows. The coordinate system is made out of all vibrational degrees of freedom and consists of a hyper-radius $\rho$ and a set of hyper-angles $\Omega$. By the construction, the hyper-radius describes uniformly all dissociation channels and, therefore, can be viewed as a generalized dissociation coordinate. One can think of the hyper-radius as related to the size of the molecular system, whereas the hyper-angles {as related to its shape}. Once the coordinate system is set, the ionic target vibrational Hamiltonian is diagonalized in the hyper-angles space for several values of $\rho$ yielding eigenvalues that represent the hyperspherical adiabatic potential energies $U_a(\rho)$ of the ion with $a=1,2\cdots$ labeling the different eigenvalues. With the potential energies in hands, the DR and VE of the polyatomic ion can considered using all the techniques developed for diatomic ions by replacing the interatomic distance $R$ with the hyper-radius $\rho$. The main difference is a much larger number of (hyperspherical adiabatic) channels compared to a typical situation of diatomic ions where only one or a few electronic channels of the target are taken into account as, for example, demonstrated above for the CH$^+$ ion.

The third approach is based on the normal mode approximation for the vibrational manifold \cite{jungen08a,jungen08b,jungen09,jungen2010low}, which usually provides a good description of the vibrational dynamics of molecular systems near the equilibrium geometry. Because, in general, normal coordinates are easy to determine for small polyatomic ions, they are perfectly suitable for determination of rovibrational excitation cross sections. However, due to their lack of dissociation limit, they cannot represent the dissociative dynamics that are needed to treat the DR process. Thus, additional steps are needed if one wants to use normal modes to treat the vibrational dynamics of DR, and they are described below.

\subsubsection{Normal mode approach for DR and VE}

If one is interested in thermal rate coefficients or cross sections with a relatively low energy resolution, one important observation is that resonances associated with closed rotational or vibrational channels are smeared out. In fact, at present, the highest resolution achieved in storage ring experiments measuring DR cross sections is a few cm$^{-1}$. Then, one has to add to this uncertainty a different one associated with the toroidal sections of the storage-ring or merged-beam set ups, where relative velocities of ions and electrons are not perfectly matched. In this situation, the energy splitting between two Rydberg resonances, which represent closed {vibrational} channels for autoionization {and increase the probability for dissociation in the DR process,} is smaller than the experimental resolution. {It is therefore} reasonable to evaluate the DR cross sections averaged over the energy splitting {as it is made by Mikhailov {\it et al.}\cite{mikhailov06}. Such a DR cross section averaged over the energy interval between three consecutive  autoionizing vibrational resonances, having with energies $\epsilon_n$, $\epsilon_{n+1}$, and $\epsilon_{n+2}$ and responsible for a temporary capture of the electron by the target ion, is 
 \begin{equation}
  \langle\sigma(E_{el})\rangle=\frac{1}{\Delta_n+\Delta_{n+1}}\int_{\epsilon_n-\Delta_n}^{\epsilon_n+\Delta_{n+1}}
   \sigma(E'){\rm d}E'\,, \quad 
   \Delta_{n+1}=\frac{\epsilon_{n+1}-\epsilon_n}2\,.
 \end{equation}}
As shown in  Ref. \cite{mikhailov06}, {if autoionization lifetimes of the resonances are large compared to their predissociation lifetimes, the} averaging procedure gives the following expression for the DR cross section
\begin{equation}
\label{eq:cs0}
\sigma(E_{el})=\frac{2a_0\pi^2}{e^2 k^2}
\Gamma_{v}
\nu^3\,,
\end{equation}
where $E_{el}=(\hbar k)^2/(2m_e)$, $k$, $e$, and $m_e$ are the kinetic energy, wave number, charge and mass of the incident electron respectively, $a_0$ is the Bohr radius,  $\Gamma_{v}$ is the width of the resonance produced by the closed vibrational channel $v$ of the ion, and $\nu$  the effective quantum number of the resonance with respect to the closed channel.

Provided that one neglects the presence of perturbing resonances associated with other vibrational or electronic channels, the widths $\Gamma_{v}$ scale with the effective quantum number of the Rydberg electron $\nu$ as $1/\nu^3$ and, therefore the product $\Gamma_{v}\nu^3$ is energy-independent. As shown in Refs.~\cite{mikhailov06,douguet09,samantha14}, the product $\Gamma_{v}\nu^3$ is related to non-diagonal matrix elements of the reactance $\langle v,\Lambda|\hat {\cal K}|v',\Lambda'\rangle$ or scattering $\langle v,\Lambda|\hat S|v',\Lambda'\rangle$ matrices, where the indices $v',\Lambda'$ refer to the initial vibrational and electronic states of the target ion and the indices $v,\Lambda$ correspond to the vibrational and electronic states producing the Rydberg series of resonances  $\Gamma_{v}$. Expressed in terms of the scattering-matrix element, the averaged cross section  can be written as \cite{douguet09,samantha14}
\begin{equation}
\label{eq:cs}
\sigma=\frac{\pi}{k^2}\vert\langle v,\Lambda|\hat S|v',\Lambda'\rangle\vert^2\,.
\end{equation}
The scattering matrix element is computed as the integral \cite{atabek74}
\begin{equation}
\label{eq:vib_FT}
\langle v,\Lambda|\hat S|v',\Lambda'\rangle=\int d{\cal Q}\langle v|{\cal Q}\rangle S_{\Lambda,\Lambda'}({\cal Q})\langle{\cal Q} |v'\rangle\,.
\end{equation}
in which the symbol ${\cal Q}$ refers collectively to all normal coordinates labeled as $q_i$ (${\cal Q}=\{q_1,q_2,\cdots\}$), describing the vibration of the ion. Using the relationship between the scattering matrix $\hat S$ matrix and the quantum defect matrix $\hat\mu$, $\hat S=\exp(2\pi i \hat\mu)$, and expanding $\mu_{\Lambda,\Lambda'}({\cal Q})$ in a Taylor series around the equilibrium configuration ${\cal Q}_0$  of the ion
\begin{equation}
\label{eq:taylor}
\mu_{\Lambda,\Lambda'}({\cal Q})=\mu_{\Lambda,\Lambda'}({\cal Q}_0)+\sum_i \frac{\partial \mu_{\Lambda,\Lambda'}}{\partial q_i}q_i+\dots
\end{equation}
allows to express analytically the matrix element of the quantum defect as long as the derivatives in Eq.~(\ref{eq:taylor}) are known. 

It is convenient to use dimensionless coordinates $q_1, q_2,\cdots$, which are related to the length-unit normal coordinates $S_1, S_2,\cdots$ as $q_i=S_i\sqrt{\mu_{red}\omega/\hbar}$, where $\mu_{red}$ and $\omega$ are the reduced mass and the frequency of the normal mode\cite{davydov76}. Assuming that the ion is initially in its ground vibrational state $\vert v'\rangle= \vert 0\rangle$ and retaining only zero- and first-order terms in the above expansion, Eq. (\ref{eq:cs}) takes the form \cite{kokoouline11a}
\begin{eqnarray}
\label{eq:cs2}
 \sigma_i(E_{el})=\frac{4\pi^3}{k^2}\left({\frac{\partial \mu_{\Lambda,\Lambda'}}{\partial q_i}}\right)^2\vert\langle v_i|\hat q_i|0\rangle\vert^2\,.
\end{eqnarray}
The index $i$ in $v_i$ and $\sigma_i$ is used to stress that the capture occurs into the $q_i$ vibrational mode excited by one vibrational quantum $v_i$. It is more convenient to use the effective quantum numbers $\nu({\cal Q})=n-\mu({\cal Q})$, where $n$ is the principal quantum number, rather than quantum defects.  In the harmonic oscillator approximation, the matrix element $\langle v_i|\hat q_i|0\rangle=\delta_{v_i,1}/\sqrt{2}$. This gives
\begin{eqnarray}
\label{eq:cs3}
 \sigma_i(E_{el})=\frac{2\pi^3}{k^2}\left({\frac{\partial \nu_{\Lambda,\Lambda'}}{\partial q_i}}\right)^2 \theta(\hbar\omega_i-E_{el})g\delta_{v_i,1}\,.
\end{eqnarray}
In the above equation the spin degeneracy factor $g$ is explicitly specified.

The equation above describes in fact the process of electron capture into vibrational resonances associated with closed vibrational levels. It gives the DR cross section only if the autoionization lifetime of these resonances is much larger than the dissociation lifetime, which has been verified for the DR process of H$_3^+$ \cite{kokoouline05b}. This assumption is believed {to} be valid also for other polyatomic molecules because the neutral molecule formed after the electron capture goes quickly to geometries in which autoionization is forbidden due to the repulsive character of the potential energy surfaces of the neutral molecule near the geometry of equilibrium of the ion. 

{Due to the procedure of averaging over the interval of electron energies corresponding to the energy splitting between vibrational autoionizing resonances, discussed above, the theoretical DR cross section at low energies is featureless and behave simply as 1/$E_{el}$. Due to a relatively low resolution in all DR experiments with polyatomic ions, except a few experiments with H$_3^+$, the experimental resolution is too low to resolve individual Rydberg resonances in the DR spectra. Therefore, the averaging procedure is justified. The interval of applicability of the procedure is determined by the approximation of the energy-independence of the product $\Gamma_v\nu^3$ in Eq.~(\ref{eq:cs0}). The product varies to the same degree as strongest couplings between different partial waves in the scattering matrix. In the absense of electronic resonances in the spectra (such resonances can influence the process for closed-shell ions at relatively high energies, above a few eV), such a variation is weak and can be neglected for the interval of electron energies up to 1 eV. The uncertainty introduced by the averaging procedure is significantly smaller than the uncertainty due to the quantum-defect approach discussed in this section.}

\begin{figure}[t]
\includegraphics[width=0.75\linewidth]{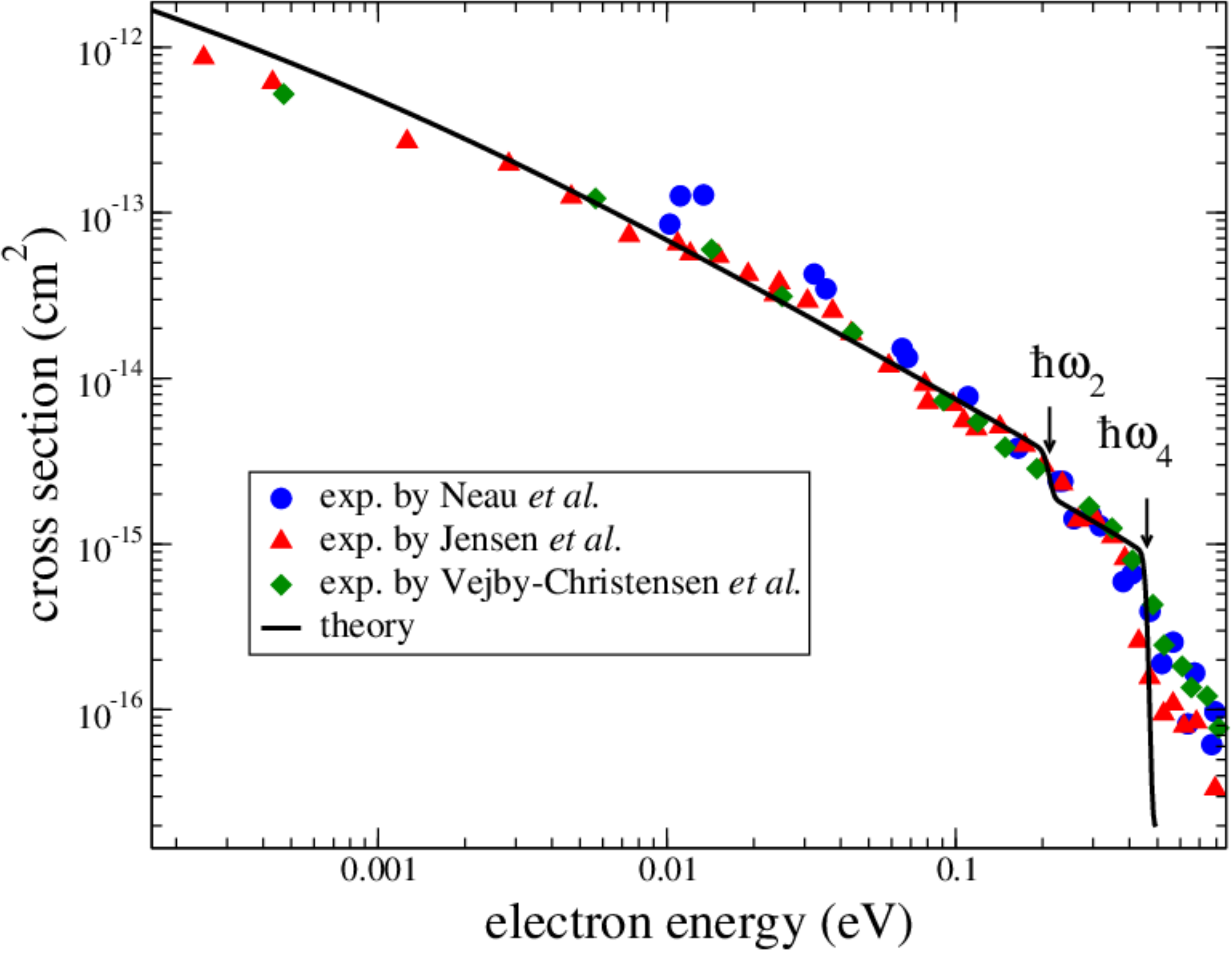}
\caption{\label{fig:cs_H3O} Comparison of theoretical \cite{douguet12a} and experimental \cite{christensen97} cross sections for DR of H$_3$O$^+$. The experimental data are obtained in the ASTRID storage ring, the theoretical cross section is obtained using Eqs.~(\ref{eq:cs3}) and (\ref{eq:Hint_JT}) and applying a convolution with the experimental distribution over collision energies.  The arrows indicate energies of excitation of a quantum of the  $q_2$ and $q_4$  degenerate modes.
}
\end{figure}

In order to apply Eq.~(\ref{eq:cs3}) one should know the behavior of the coupling elements in the matrix $\hat\nu$ of the principal quantum number. In practice, the principal quantum numbers are obtained in bound-state calculations of highly excited electronic Rydberg states. The principal quantum numbers obtained in this way correspond to eigenvalues of the matrix $\hat\nu$, i.e. they don't provide the information about the couplings between closed electronic states. In order to obtain the couplings, model Hamiltonians are developed for different types of the molecular ions. The form of a particular model Hamiltonian depends on the symmetry of the ion. Based on the model Hamiltonian, a model for the matrix of principal quantum number is obtained and its elements $ \nu_{\Lambda,\Lambda'}$ are computed fitting the eigenvalues of matrix evaluated in bound-state calculations to the model matrix. This approach was used in several studies \cite{mikhailov06,douguet08b,douguet09,douguet11a,douguet12a}. For example, for the H$_3$O$^+$ and CH$_3^+$ ions, the model Hamiltonian is constructed \cite{douguet12a} taking into account the Jahn-Teller coupling between degenerate electronic states and vibrational modes. Based on the model Hamiltonian, the model matrix of  principal quantum number for H$_3$O$^+$ has the following form
\begin{equation}
\hat\nu(\rho,\varphi)=\left(
\begin{array}{ccc}
\nu^{(0)}_e &\kappa_{ea}\rho e^{i\varphi } &\kappa_{ee}\rho e^{-i\varphi } \\
\kappa_{ea}\rho e^{-i\varphi }& \nu^{(0)}_{a} & \kappa_{ea}\rho e^{i\varphi }\\
\kappa_{ee}\rho e^{i\varphi }& \kappa_{ea}\rho e^{-i\varphi }& \nu^{(0)}_e
\end{array}
\right) \,,
\label{eq:Hint_JT}
\end{equation}
in the basis of three electronic states taken into account: an $a_1$ and the doubly-degenerate $e$ states.  The coordinates  $\rho$ and $\varphi$ are polar versions of the $x$- and $y$- components of the $q_2$ or $q_4$ degenerate modes of the ion. The parameters $\kappa_{ee}$ and $\kappa_{ea}$ are real, $\rho$- and $\varphi$-independent, and obtained by a fit to the {\it ab initio} results: The eigenvalues of the above matrix should be equal to the values obtained from {\it ab initio} bound-state energies. More details about the construction of the matrix (\ref{eq:Hint_JT}) can be found in Ref.~\cite{douguet12a}.

\begin{figure}[t]
\includegraphics[width=0.75\linewidth]{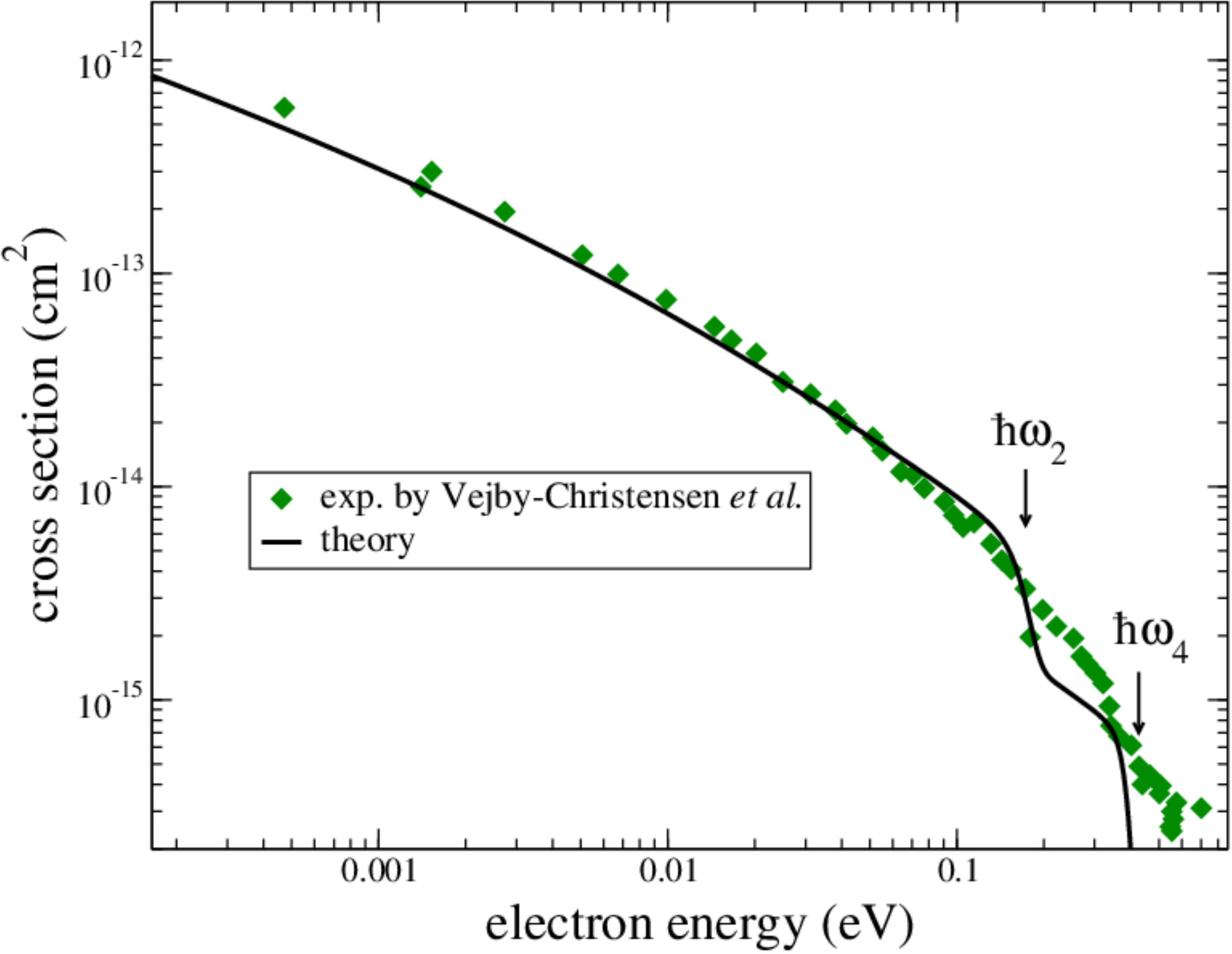}
\caption{\label{fig:cs_CH3} Same as Fig.~\ref{fig:cs_H3O}, but for the DR in CH$_3^+$.
}
\end{figure}

The figures  \ref{fig:cs_H3O}  and \ref{fig:cs_CH3} show the DR cross sections for the H$_3$O$^+$ and  CH$_3^+$ ions obtained using the approach described above. The theoretical cross sections agree well with the experimental data from the storage-ring experiments. A more detailed discussion of the theoretical approach and the comparison with the experiment can be found in Ref. \cite{douguet12a}.

\subsection{Beyond quantum defect approach}

The theoretical approach described above requires couplings in the matrices of quantum defect or principal quantum number. The couplings are evaluated assuming a model for the Hamiltonian of interaction between electronic and vibrational degrees of freedom of the ion+electron system. The approach was successful for $C_{3v}$, $D_{3h}$, and $T_d$ \cite{douguet12b} molecular ions, where the Jahn-Teller coupling has the largest contribution to the DR  and vibrational excitation cross section. This means that the model Hamiltonian and, correspondingly, the model couplings in the matrix of Eq.~(\ref{eq:Hint_JT}) are able to represent entirely the major physics of the DR and of the VE processes. 

Trying to apply similar ideas to other types of polyatomic ions, such as linear ions of the $C_{\infty v}$ symmetry group, was not as successful as it was for the $C_{3v}$, $D_{3h}$, and $T_d$ ions. Initially, it was suggested that the Renner-Teller coupling could be the main mechanism for the DR process in the $C_{\infty v}$ ions. The idea was developed \cite{mikhailov06,douguet08b,douguet09} on the example of the HCO$^+$ ion. The model Hamiltonian and the couplings in the matrix of principal quantum number were obtaining based on the Renner-Teller coupling mechanism. However, the obtained theoretical DR cross section was significantly smaller, {by} a factor of $2-10$, than all available experimental data. In addition, the fitting procedure of bound state energies to the model matrix of quantum defects is not unique. After many efforts it was concluded that the Renner-Teller interaction is not the only one that contributes to the DR process. The vibronic coupling between electronic states other than involved in the Renner-Teller interaction, should be accounted for to describe the DR process in linear ions.

\begin{figure}[t]
\includegraphics[width=0.75\linewidth]{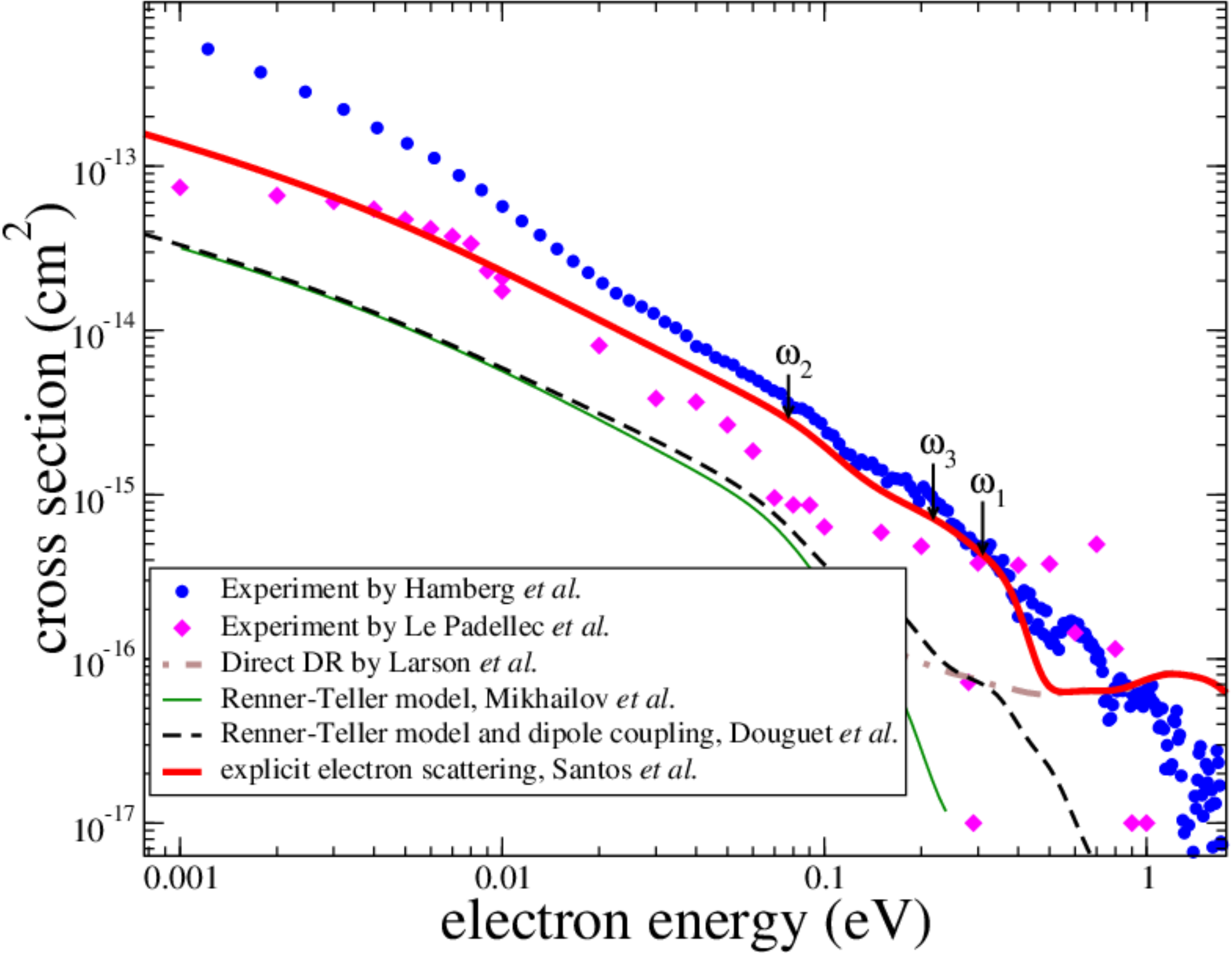}
\caption{\label{fig:cs_HCO} Comparison of different theoretical and experimental DR cross sections for HCO$^+$. The red solid line is the theoretical result \cite{samantha14} obtained using Eq.~(\ref{eq:cs_DR}). The theoretical results obtained using the model Renner-Teller Hamiltonian and the approach based on quantum defects (discussed above) are shown with green \cite{mikhailov06} and dashed black \cite{douguet08b} curves. The experimental results from a merged beam experiment \cite{lepadellec97} and CRYRING storage ring \cite{hamberg2014experimental} are shown by purple and blue dots, respectively. The first vibrational thresholds of each normal mode of HCO$^+$  are indicated by arrows.
}
\end{figure}

The difficulties with the above theoretical approach and the availability of electron-scattering codes suggested that one can use the results of electron-scattering calculations directly, without employing model Hamiltonians and quantum defect coupling models: Instead of computing energies of excited electronic states of the neutral molecule, one can simply use scattering or reaction matrices obtained from electron-scattering codes, such as the UK R-matrix \cite{tennyson2010} or complex Kohn method \cite{mccurdy89,orel91}. Therefore, if one expands the scattering matrix in Eq.~(\ref{eq:vib_FT}) in a way similar to Eq.~(\ref{eq:taylor}), one obtains for the DR  cross section 
\begin{eqnarray}
\label{eq:cs_DR}
\sigma^{DR}(E_{el})=\frac{\pi\hbar^2}{4 m_e E_{el}}\sum_{i=1}^3 g_i\sum_{ll'\lambda\lambda' }\left| \frac{\partial S_{l\lambda,l'\lambda'}}{\partial {q}_i}\right|^2\theta(\hbar\omega_i-E_{el})\,.
 \end{eqnarray}
Here $i$ runs over all modes of the target molecules, $g_i$ is the degeneracy factor for the mode $i$, $l,\lambda$ and $l',\lambda'$ are the indices of the partial waves and their projections on a chosen quantization axis in the molecular reference frame. The above formula gives the DR cross section assuming that the target ion is on its ground vibrational level.

The cross section for vibrational excitation of the mode $i$ from $v_i=0$ to $v_i=1$ is \cite{ayouz2016,yuen2019dissociative} 
\begin{eqnarray}
\label{eq:cs_VE}
\sigma^{VE}_i(E_{el})=\frac{\pi\hbar^2}{4 m E_{el}}g_i\sum_{ll'\lambda\lambda' }\left| \frac{\partial S_{l\lambda,l'\lambda'}}{\partial {q}_i}\right|^2 \theta(E_{el}-\hbar\omega_i)\,.
 \end{eqnarray}

In practice \cite{ayouz2016,slava2018,yuen2019dissociative}, if one uses the UK R-matrices codes \cite{tennyson2010}, the scattering matrix should be obtained from the reactance matrix $\hat K$ as 
\begin{eqnarray}
\label{eq:K2S}
\hat S =\frac{\hat 1 + i \hat K}{\hat 1 - i \hat K}\,,
\end{eqnarray}
where $\hat 1$ is the identity matrix, and it has to be computed  for {\it two values only} of each normal coordinates $q_i$, such that the derivative  $\partial S_{l\lambda,l'\lambda'}/\partial q_i$ of the scattering matrix with respect to the normal coordinate $q_i$ could be computed by the simple finite-difference formula. When the complex Kohn method  is employed \cite{samantha14}, electron scattering calculations produce directly the matrix $\hat S$ and, therefore, the step of Eq.~(\ref{eq:K2S}) is not needed.

Figure \ref{fig:cs_HCO} compares the available experimental data from two experiments with  theoretical results obtained using the quantum defect approach and explicit scattering calculations. The data from the two experiments differ quite significantly at low electron energies. As discussed in Ref. \cite{samantha14}, in the storage-ring experiment, the two stable isomers of HCO$^+$  (HCO$^+$ and HOC$^+$) could be present. The theoretical DR cross section for HOC$^+$ is significantly larger than the one for HCO$^+$ \cite{samantha14}. Therefore, the differences in the experimental data \cite{lepadellec97,hamberg2014experimental} could be explained by {a larger fraction of} HOC$^+$ {in the CRYRING storage ring \cite{hamberg2014experimental} compared to the} merged beam experiment \cite{lepadellec97}. {The merged beam cross section agrees quite well with the pure-HCO$^+$ theoretical result and the storage-ring data \cite{hamberg2014experimental} could be reproduced in calculations if one assumes that some fraction of ions in the experiment is the form of the HOC$^+$ isomer, which might mean that in the CRYRING experiment the two isomers are present.} The cross section obtained using the quantum defect approach, shown with green \cite{mikhailov06} and dashed black \cite{douguet08b} curves in the figure, underestimate the DR cross section. As discussed in Ref. \cite{samantha14}, this is due to the vibronic coupling between symmetric and asymmetric stretching modes neglected in the quantum defect approach used in previous theoretical studies \cite{mikhailov06,douguet08b}. 

\begin{figure}[t]
\includegraphics[width=0.75\linewidth]{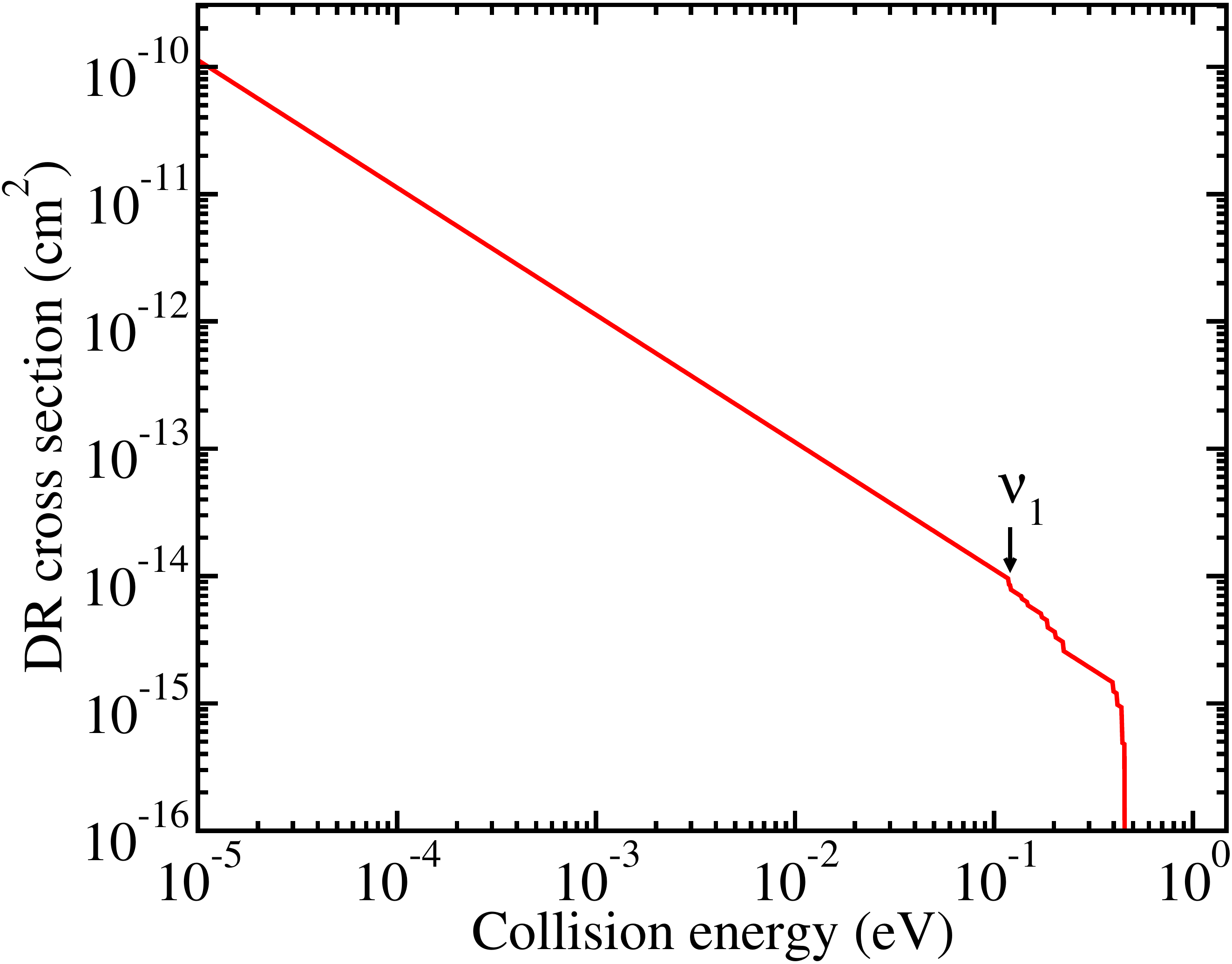}
\caption{\label{fig:cs_CH2NH2} Dissociative recombination of ground state CH$_2$NH$_2^+$: cross section. 
}
\end{figure}

The only COM target ever involved in a DR and VE theoretical study published so far is CH$_2$NH$_2^+$ \cite{yuen2019dissociative}. This cation can act as a precursor of simple amino acids. {We have applied the approach discussed above to this ion.} Figure \ref{fig:cs_CH2NH2} shows the DR cross section for the target in its vibrationally ground state as a function of collision energy. For energies higher than $0.1$ eV, the cross section drops in a stepwise manner because the scattering electron excites the vibrational level of the ionic target by one quanta. 

\begin{figure}[t]
\includegraphics[width=0.75\linewidth]{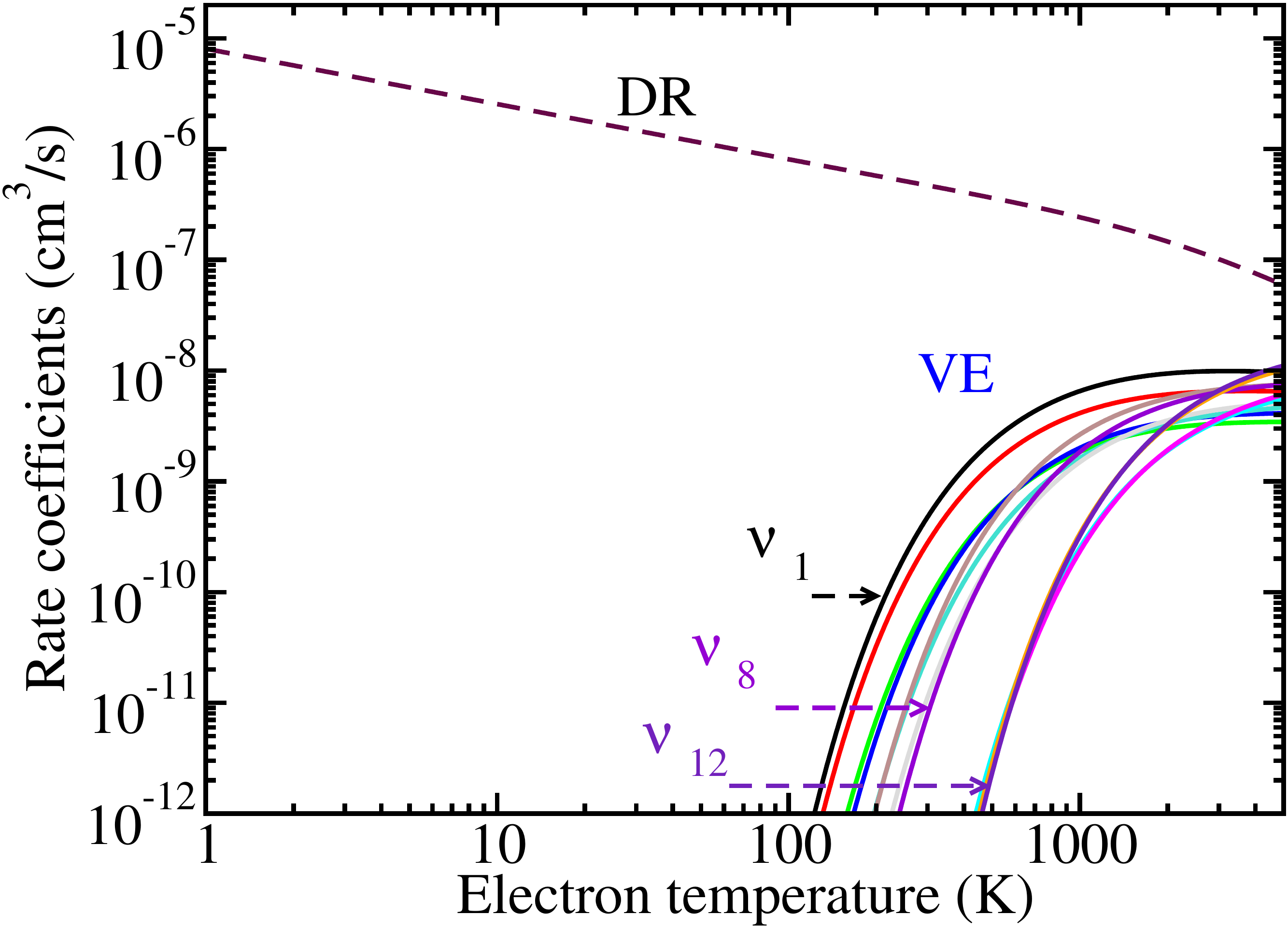}
\caption{\label{fig:rate_CH2NH2}  Dissociative Recombination  (DR, dashed line) and Vibrational Excitation (VE, solid lines) of CH$_2$NH$_2^+$ in its ground state. 
VE rate coefficients are arranged from low to high frequencies. The vibrational modes 1, 8 and 12 are indicated with arrows for visual guidance. }
\end{figure}

Figure \ref{fig:rate_CH2NH2} displays the thermally averaged VE and DR rate coefficients versus the electron temperature for all the normal mode frequencies. For $T <$ 400 K, the DR and VE rate coefficient goes as $(k_b T)^{-1/2}$ and $(k_b T)^{-1/2} \exp{\left( -\hbar \omega_i/k_b T\right)}$ respectively. At higher temperature, as vibrational excitation is more probable, the DR rate coefficient decreases faster than $(k_b T)^{-1/2}$.

{To the best of our knowledge, there is no experimental data, cross sections or rate coefficients, on DR or VE processes in electron-CH$_2$NH$_2^+$ collisions. It would desirable if, for example, DR rate coefficients are measured in a storage-ring or merged-beam experiments. Such a measurement would help to validate the applicability of the presented theoretical approach for ions with more than 4 atoms.} 

\section{Conclusions}

The Multichannel Quantum Defect Theory allows the realistic modeling of the reactive collisions between electrons and molecular cations, provided that the molecular structure of the target and of the neutral complex has been explored
and quantitatively characterized by quantum chemistry and R-matrix methods.

High accuracy has been achieved for some of the diatomic species, and the progressive account of the numerous mechanisms, and interactions - Rydberg-valence and Rydberg-Rydberg couplings, rotational and core-excited effects - results in theoretical cross sections and rate coefficients in increasing agreement with the measurements in storage rings. {However, so far,  with the exception  of rough estimations for CO$^+$ \cite{mezei2015,moulane2018}, we did not produce branching ratios for the systems studied in this work. The major reason for this is the lack of data for the interactions at large internuclear distances, which may change the dissociation dynamics with respect to the tendencies induced at small distances. This important issue is subject of further studies.}

As for the polyatomic ions, we would like to stress that after several iterations and several years of development, a reliable and relatively simple theoretical approach was developed to study the dissociative recombination and rovibrational excitation of small molecular ions having a closed electronic shell. The method is based on explicit electron-molecule scattering calculations, which could be performed using different existing codes and methods. The approach was successfully applied for a variety of molecular ions, the largest ones so far being CH$_2$NH$_2^+$ \cite{yuen2019dissociative} and NH$_2$CH$_2$O$^+$ \cite{yuen2019_NH2CH2O}. {The theoretical studies of the DR process in the CH$_2$NH$_2^+$ \cite{yuen2019dissociative} and NH$_2$CH$_2$O$^+$ \cite{yuen2019_NH2CH2O} ions and their importance in the chain of prebiotic reactions in the interstellar clouds invite experimentalists to perform measurements of the DR rate coefficients. Such experimental measurement will validate the theoretical approach or will prompt a development of a better theoretical method for middle-size polyatomic ions.}

These developments will contribute in the near future to a better understanding of the role of the collisions of electrons with molecular cations in the formation of the complex organic molecules.

\begin{acknowledgments}
JZsM is grateful for the support of the National Research, Development and Innovation Fund of Hungary, under the K18 funding scheme with project no. K 128621. IFS acknowledges support from 
Agence Nationale de la Recherche via the project MONA,  
from the CNRS via the  GdR  TheMS, 
from La R\'egion Normandie, FEDER and LabEx EMC$^3$ via  the projects Bioengine, PicoLIBS, EMoPlaF and CO$_2$-VIRIDIS, and from the Programme National 'Physique et
Chimie du Milieu Interstellaire' (PCMI) of CNRS/INSU with INC/INP co-funded
by CEA and CNES. This work was  supported by the National Science Foundation, Grant No. PHY-1806915.
\end{acknowledgments}

section*{References}
\bibliography{achemso-jzsmezei,DR}

\end{document}